\documentclass[11pt,preprint]{aastex}

\newcounter{romnum}

\begin{document}

\title{A Magnetic $\alpha\omega$ Dynamo in Active Galactic Nuclei
Disks: I. The Hydrodynamics of Star-Disk Collisions and Keplerian
Flow}

\author{Vladimir I. Pariev\altaffilmark{1}\altaffilmark{2} and Stirling
A. Colgate}
\affil{Theoretical Astrophysics Group, T-6, Los Alamos National
Laboratory, Los Alamos, NM 87545}
\altaffiltext{1}{Lebedev Physical Institute, Leninsky Prospect 53,
Moscow 119991, Russia}
\altaffiltext{2}{Currently at Physics Department, University of
Wisconsin-Madison,  1150 University Ave., Madison, WI 53706}

\begin{abstract} A magnetic field dynamo in the inner regions of the
accretion disk surrounding the supermassive black holes in Active
Galactic Nuclei (AGNs) may be the mechanism for the generation of
magnetic fields in galaxies and in extragalactic space. We argue that
the two  coherent motions produced by 1) the Keplerian motion and 2)
star-disk collisions, numerous  in the inner region of AGN accretion
disks, are both basic to the formation of a robust, coherent dynamo
and consequently the generation of large scale magnetic fields. 
In addition we find that the predicted rate, 10 to 100 per year
at $\sim 1000 r_g$, $r_g$ the gravitational radius,  and the
consequences of star-disk collisions are qualitatively, at least,
not inconsistent  with observations of broad emission and absorption
lines. They are frequent enough to account for an integrated dynamo
gain, $e^{10^{9}}$ at $100 r_g$,  many orders of magnitude greater
than required to amplify any seed field no matter how small. The
existence of extra-galactic, coherent, large scale magnetic fields
whose energies greatly exceed all but massive black hole energies is
recognized.  In paper II \citep{pariev06} we argue that in order to produce a dynamo
that can access the free energy of black hole formation and produce
all the magnetic flux in a coherent fashion the existence of these
two coherent motions in a conducting fluid is required. The
differential winding of Keplerian motion is obvious, but the disk
structure depends upon the model of "$\alpha$", the transport
coefficient of angular momentum chosen. The counter rotation of
driven plumes in a rotating frame is less well known, but
fortunately the magnetic effect is independent of the disk model. 
Both motions are discussed in this paper, paper~I.  The description
of the two  motions are preliminary to two theoretical derivations
and one numerical simulation of the $\alpha\omega$ dynamo in paper
II.

\end{abstract}

\section{Introduction}\label{sec_intro}

The need for a magnetic dynamo to produce and amplify the immense
magnetic fields observed external to galaxies and in clusters of
galaxies has long been recognized. The theory of kinematic magnetic
dynamos has had a long history and is a well developed subject by
now. There are numerous monographs and review articles devoted to the
magnetic dynamos in astrophysics, some of which are:
\citet{parker79};
\citet{moffatt78};
\citet{stix75}; \citet{cowling81}; \citet{roberts92};
\citet{childress90}; \citet{zeldovich83};
\citet{priest82};
\citet{busse91}; \citet{krause80}; \citet{biskamp93};
\citet{mestel99}. Hundreds of  papers on magnetic dynamos are
published each year. Three main astrophysical areas, in which
dynamos  are involved, are the generation of magnetic fields in the
convective zones of planets and stars, in differentially rotating
spiral galaxies, and in the accretion disks around compact objects. 
The possibility of production of magnetic fields in  the central
parts of the black hole accretion disks in AGN has been  pointed out
by
\citet{chakrabarti94} and the need and possibility for a robust
dynamo by \citet{colgate97}. Dynamos have been also
observed in the laboratory in the Riga experiment \citep{gailitis00,
gailitis01} and in Karlsruhe experiment \citep{stieglitz01}. 
Recently,  counter
rotating, opposed jets  or
open-flow geometries, such as the the Von K{\'a}rm{\'a}n Sodium
(VKS) Experiment and the Madison Dynamo Experiment, have been
designed to explore less constrained flows ~\citep{bourgoin04,
spence06}. So far, neither of these experiments have reported
sustained magnetic field generation despite predictions of positive
gain in laminar flow theory and calculations.  The null result has
been ascribed to the deleterious effects of enhanced turbulent
diffusion of large-scale turbulence.

\subsection{The Need for a Robust Astrophysical Dynamo}
\label{subsec_intro1}

Why, with all the thousands of research papers, very many successes,
and even experimental verification of dynamo theory 
in constrained flows 
is there a need
for a new paradigm for the generation of intergalactic scale
astrophysical magnetic fields?  We claim that  the plume-driven
$\alpha\omega$ dynamo in the black hole accretion disk is a unique
solution to the need for the largest dynamos of the universe,
because the flow is naturally constrained by a gradient in angular
momentum and by the transient dynamical behavior of plumes
in contrast to the large turbulence of unconstrained flows.
(A discussion of the role of convective plume-driven $\alpha\omega$
dynamos in stars will be reserved for another paper, because the
mechanism of the production of large scale plumes in the convective
zone of stars is radically different from the production of plumes
by high velocity stars plunging frequently through the accretion
disk.)

The minimum energy inferred from radio emission observations of
structures or so-called radio lobes within clusters and external to 
clusters by both synchrotron emission and Faraday rotation
\citep{kronberg94, kronberg01} are so large,
$\sim 10^{59}\,\mbox{ergs}$ and and up to  $\sim
10^{61}\,\mbox{ergs}$ respectively, compared to galactic energies
in fields $10^{-7}$  as large  and gravitational binding energies
$10^{-3}$ as large, that only  the energy of formation of the
central massive black hole (hereafter, CMBH) of every galaxy in its
AGN phase,  $\sim 10^{62}\,\mbox{ergs}$, becomes the most feasible
astrophysical known source of so much energy. This statement is based
upon the recognition that $\sim 10^8$  neutron stars have been
created in the galaxy in a Hubble time, or only 
$\sim 10^6$ in the life time of a radio lobe of $\sim 10^8\,\mbox{yr}$.  Each
supernova results in
$10^{51}\,\mbox{ergs}$ of kinetic energy, the rest being emitted in neutrinos
so that $\sim 10^{57}\,\mbox{ergs}$ of kinetic energy becomes available for the
production of magnetic energy within the necessary time. If one
considers the difficulty of summing the magnetic field from many, presumably
incoherent sources, the likelihood of many stellar sources
contributing to the coherent field of radio lobes seems remote.

In order to access this energy of formation the conversion of
kinetic to magnetic energy is required. This in turn requires a
mechanism to multiply or exponentiate  an initial field  up to the
back reaction limit.  
This limit is where  the Ampere
force does a large  work to significantly alter the accretion motion
thus converting the  kinetic energy to magnetic energy.   
Because the specific angular momentum of
matter accreted onto the CMBH is $\sim 10^3\, \mbox{to}\, 10^4$
greater than possible for accretion at  $r_g$, the result is the
universal Keplerian motion of an accretion disk and so the access of
this free energy must be in the form of a back reaction torque that
transforms kinetic to magnetic energy.

A robust dynamo is one that can potentially convert a large
fraction of the available  mechanical energy or free energy of the
accretion disk into magnetic energy. 
A further advantage of the $\alpha\omega$ 
dynamo in the CMBH accretion disk is that the exponential gain within
$100$ gravitational radii of the CMBH is so large, some fraction $f$
per turn,  or gain $= e^{fN}$, $N
\sim 10^{9}$ turns in the
$10^8$  years of formation, that the origin and strength  of the
initial (seed) field becomes moot.

\subsection{The Robust $\alpha\omega$ Dynamo}
\label{subsec_intro2}
Such a dynamo  has conceptually become feasible because of the
recognition of a relatively new, coherent, large scale, robust
source of helicity.  Helicity generation, in the sense of the
$\alpha\omega$ dynamo, is the driven deformation of the conducting
fluid that converts an amplified (by differential winding) toroidal
field back into the initial, (radial), poloidal field. In our case it
is caused by the rotation of driven, diverging plumes in a rotating
frame \citep{beckley03, mestel99, colgate99}. The advantage of
driven plumes as a source  of helicity as compared to turbulent
motions within the disk is that the flow displaces fluid and
entrapped flux well above the disk, several scale heights, and then
rotates the flux on average a quarter turn before merging again into
the disk.  Such an ideal deformation is then a large coherent 
(single direction) source of helicity. These plumes are presumably
driven by many stars in orbits repeatedly plunging through the disk,
but comprising only a small mass fraction, $\sim 10^{-3}$, of the
matter accreted to form and grow the CMBH to $\sim 10^8 \,M_{\sun}$.
The twisting  or relative rotation of the plumes occurs because of 
partially conserved angular momentum of the plume itself as  its
moment of inertia increases due to its expansion or divergence while
progressing in height.  The repeatable fractional turn before
merging with the disk occurs because the cooling plume matter falls
back to the disk in half a turn of the disk. This translation and
rotation  twists  the embedded toroidal magnetic field. Furthermore,
the angle of twist is in the same direction for all plumes, opposite
to the rotation of the disk, and furthermore the angle of this twist
is  limited to
$\Delta \phi \simeq  - \pi /2$ radians of rotation for each
occurrence.  This nearly ideal repetitive driven deformation leads
to a robust dynamo, one where both motions are not likely to  be
easily damped by back reaction except at the full Keplerian stress.
Such a dynamo is not dependent upon a net helicity derived from 
random turbulent motions. The limitation of turbulently derived
helicity due to early back reaction is discussed later, but first we
discuss the preference for a finite angle, specifically $(2n +1)
\pi /4$ angle of rotation in $n$ periods of rotation for an
effective  helicity. (Preferably  $n = 0$.)

\subsection{The Original $\alpha\omega$ Dynamo}
\label{subsec_intro3}

The original proposal of \citet{parker55, parker79} of the
$\alpha\omega$ dynamo in rotationally sheared conducting flows,
seemed to be the logical answer to the problem of creating the large,
highly organized fields of stars and  galaxies as revealed by 
polarized synchrotron emission and Faraday rotation maps. Here the
radial component of a poloidal field is wrapped up by differential
rotation into a much stronger toroidal field. Then as proposed by
Parker, cyclonic motions of geostrophic flow twist and displace
axially a fraction of the toroidal flux back into the poloidal
direction. Subsequent merger of this  small component of poloidal
flux with the large scale original poloidal flux by resistive
diffusion or reconnection completed the cycle. The later process of
merging the small scales to create the large scales is referred to as
mean field dynamo theory. There were two apparently insurmountable
problems with this theory. The first, as argued by
\citet{moffatt78} and as discussed in
\citet{roberts92} was that  geostrophic cyclonic flows, with
negative pressure on axis, make very many revolutions before
dissipating therefore reconnecting the flux in an arbitrary
orientation.  Hence, the orientation of any newly formed  component
of poloidal  flux would be averaged to near zero. The star-disk
driven plumes, on the other hand, avoid this difficulty by falling
back to the disk in less than $\pi$ revolutions of rotation, thereby
terminating further rotation by fluid merging within the disk. The
second difficulty  was that the large dimensions of interstellar
space and finite resistivity ensured a near infinite magnetic
Reynolds number,
$\mbox{Rm} = Lv/\eta $ ($L$ the dimension, $v$ the velocity and
$\eta$ the resistivity), so that, in general, the resistive
reconnection time would become large compared to the age of the
astrophysical object. Consequently newly minted poloidal flux would
never merge with the original poloidal flux.

Currently, although the details of reconnection are poorly
understood, it is well recognized in both astrophysical
observations, theory, and in the many fusion confinement experiments
that reconnection occurs astonishingly fast, up to Alv\'en speed. As
a result, physicists concerned with the problem turned to turbulence
as the  solution, both to produce a small net helicity as well as to
produce an enhanced resistivity in order to allow reconnection of the
fluxes. Furthermore mean field theory was developed to predict the
emergence of large scale fields from the merger  of small scale
turbulent motions \citep{steenbeck66, steenbeck69a, steenbeck69b}.
Ever since, mean field turbulent dynamo theory has dominated the
subject for the last 40 years.

\subsection{The Turbulent Dynamo}
\label{subsec_intro4}

There are two principle  problems with turbulent dynamos: first,  the
difficulty of deriving  a net and sufficient  helicity from random
turbulent motions, and secondly,  the ease  with which the turbulent
motions themselves can be suppressed by the back reaction of the
field stress, in this case the multiplied toroidal field
\citep{vainshtein93}. Regardless of the source of such turbulence,
i.e., the $\alpha$ viscosity
\citep{shakura73}, the magneto-rotational instability
\citep{balbus98} or magnetic buoyancy \citep{chakrabarti94}, the
turbulent stress  will be small compared to the stress of Keplerian
motion.  The stress of the magnetic field produced will be limited by
the back reaction on this turbulence.  As discussed later the back
reaction would limit the stress of the dynamo fields  to values very
much less than the Keplerian stress.

The problem of the origin of reconnection remains, but here
turbulence in the disk can help where one needs only assume that the
flow of energy in turbulence is always dissipative and that the
fraction of magnetic energy dissipated by this turbulence may be
very small yet satisfy the necessary reconnection. Secondly, fast
reconnection (at near Alv\'en speed) in low beta, collisionless
plasmas has been modeled
\citep{li03,drake03}.

We note that we are not considering turbulence as a significant
source of helicity in the
$\alpha\omega$ dynamo, yet at the same time invoking turbulence in
order to enhance reconnection.

\subsection{The Astrophysical Consequences}
\label{subsec_intro5}

We are attempting to demonstrate that  a robust dynamo in an
accretion disk,  dependent upon a small mass fraction of orbiting
stars, becomes a dominant magnetic instability of CMBH formation. 
To the extent to which this indeed is  so and since orbiting stars and
Keplerian accretion are universal, then it becomes difficult to
avoid the conclusion that the free energy of formation of most CMBHs
would be converted into magnetic energy.

In our view the magnetic field, both energy and flux, generated by
the black hole accretion disk dynamo presumably   powers the jets
and the giant magnetized radio lobes. For us both of these
phenomena  are most likely the on-going dissipation by reconnection
and synchrotron emission of force-free helices of wound up strong
magnetic field produced by the accretion disk dynamo. (The  large
scale magnetic flux, as indicated by polarization observations where
the correlation length is of order the distance between bright knots,
M87,
\citet{owen80} is equally demanding of the coherence of the dynamo
process.) The electromagnetic mechanism of extraction of angular
momentum and  energy from the accretion disk has been proposed by
\citet{blandford76} and \citet{lovelace76}. Recently, the process of
formation of such a force-free helix by shearing of the foot-points
of the magnetic field by the rotation of the accretion disk has been
considered by \citet{lyndenbell96} and
\citet{ustyugova00}; \citet{li01a}; \citet{lovelace02}. The magnetic
dynamo in the disk is the essential part of the whole emerging
picture of the formation and functioning of AGNs, closely related to
the production of magnetic fields within galaxies, within clusters
of galaxies, and the still greater energies and fluxes in the IGM.
Black hole formation, Rossby wave torquing of the  accretion disk
\citep{lovelace99, li00, li01b, colgate03}, jet formation
\citep{li01a} and magnetic field redistribution by reconnection and
flux conversion, and finally  particle acceleration in the radio
lobes and jets are the key parts of this scenario \citep{colgate99,
colgate01}.
Finally we  note that if almost every galaxy contains a
CMBH and that if a major fraction of the free energy of its
formation is converted into magnetic energy, then only a small
fraction of this magnetic energy,  as seen in the giant radio lobes
\citep{kronberg01}, is sufficient to propose a possible feed back in
structure formation and in galaxy  formation.

\subsection{The Back Reaction Limit and Star-Disk Collisions}
\label{subsec_intro6}

The  main stream of astrophysical dynamo theory is the mean field
theory   where an exponential growth of the large scale field is
sought, while averaging over small scale motions of the conducting 
plasma usually regarded as turbulence.

The behavior of turbulent dynamos at the nonlinear stage i.e., back
reaction, when one can no longer ignore the Ampere force, is not
fully understood and is the process of active investigations
\citep{vainshtein92, vainshtein93, field99}. However, as it was
argued by
\citet{vainshtein92}, the growth  of magnetic fields as a result of
the action of the kinematic dynamo should lead to the development of
strong field filaments with the diameter of the order of
$L/\mbox{Rm}^{1/2}$, where $L$ is the characteristic size of the
system and
$\mbox{Rm}$ is the magnetic Reynolds number. The field in the
filaments reaches the equipartition value much sooner than the
large scale field, causing the suppression of the
$\alpha$ effect due to the strong Ampere force or back reaction, 
acting in the filaments. As a result, turbulent
$\alpha\omega$  dynamos may be able to account for the generation of
the large scale  magnetic fields only at the level of
$\mbox{Rm}^{-1/2}$ of the  equipartition value. Finding the
mechanism for producing and  maintaining large scale helical flows
resulting in a robust $\alpha$  effect is thus very important for
the generation of  large  scale magnetic fields of the order of the
equipartition magnitude.

One way of alleviating the
difficulty with the early quenching of the turbulent 
$\alpha$-dynamo may be a nonlinear dynamo, where the $\alpha$-effect
is maintained by the action of the large-scale magnetic field itself
rather than by a small-scale  turbulent motions. Such a nonlinear
dynamo due to the buoyancy of the magnetic  field in a rotating
medium was first proposed by \citet{moffatt78}. As magnetic flux
tubes  are rising, they expand sidewise to maintain the balance of
the pressure with the  less dense surrounding gas. This sidewise
velocity  is claimed to cause
the magnetic tube to bend  under the action of the Coriolis
force.

Calculations of the nonlinear dynamo applied to the Sun was 
performed by \citet{schmitt87} and \citet{brandenburg98}. A somewhat
different mechanism for the radial expansion of the buoyant magnetic
loops (due to the  cosmic ray pressure) was proposed in the context
of the Galactic dynamo by \citet{parker92} and detailed calculations
of the resulting mean field theory were performed by 
\citet{moss99}.  In this case the matter, cosmic rays, would not
fall back to the galaxy surface, but the inertial mass of the cosmic
rays is smaller than that of the galactic matter by $\sim 10^{-10}$,
again greatly reducing the back reaction limit. 
The buoyant dynamo can amplify the weak  large-scale magnetic
field, $B_c \sim \mbox{Rm}^{-1/2} B_{equi}$, where $B_{equi}$  is
the magnetic field in equipartition with the turbulent energy.
However, the buoyant
$\alpha$ is a fraction (generally, a small fraction) of the velocity
of the buoyant rise  of the toroidal magnetic fields, $u_B= C
(d/H)^{1/2} v_A$, where $d$ is the radius of a flux tube, $H$ is the
half thickness of the disk, $v_A$ is the Alfv\'en speed, and $C$ is a
constant of order unity. For $\mbox{Rm} \sim 10^{15}$ to
$10^{20}$ in the accretion disk, $B_c \sim 10^{-8}$ to $10^{-10}
B_{equi}$. Alfv\'en speed will be about $10^{-8}$ to $10^{-10}$ of
sound speed.  As we show below, star-disk collisions lead to a large
mass ejected above the disk and therefore result in robust, large
scale helical motions of hot gas  with the rotation velocity
exceeding the sound speed in the disk and, therefore,
$10^8$ to  $10^{10}$ times faster than the buoyant motions of the
magnetic flux tubes. Thus, 
we can safely
neglect the buoyant dynamo in  our calculations of the linear stage of
star-disk collision driven dynamo.

\subsection{Star-Disk Collisions }
\label{subsec_intro7}

It has now  been long realized that the  collisions of stars forming
the  central part of the star cluster  in AGNs with the accretion
disk  lead to the exchange and stripping (or possibly growth) of the
outer envelopes of stars and also, inevitably, a change in the 
momentum of the stars. This makes an important impact on the
dynamics  of stellar orbits. Thus the evolution of the central star
cluster may  contribute to providing accretion mass for the
formation of the CMBH  and can account for part of the observed
emission from AGNs
\citep{syer91, artymowicz93, artymowicz94, rauch95, vokrouhlicky98, 
landry98}.
\citet{zurek94} considered the physics of plasma tails produced
after star-disk collisions (see also
\citealt{zurek96}). They suggest that emission from these tails  may
account for the broad lines in quasars. Here we suggest another 
consequence of stars passing through the accretion disk, the 
generation of magnetic fields.\\

For this to happen on a large scale and at the Keplerian back
reaction limit requires multiple, repeatable coherent rotation
through a finite angle and axial translation of conducting matter
well above the disk.   We emphasize the importance of an
experimental,  laboratory demonstration of the rotation and
translation of  plumes,  driven by jets in a rotating frame
\citep{beckley03}. These laboratory plumes are the analogue of those
produced by the star disk collisions, which are  the source of the
helicity fundamental to this dynamo mechanism.

\subsection{The Structure of the Accretion Disk}
\label{subsec_intro8}

The near universally accepted view of accretion disks is that based
upon the transport of angular momentum by turbulence within the
disk.  This is the  $\alpha$-disk model, which is also referred to
as the Shakura--Sunyaev and to many is the standard model. This model
was developed by \citet{shakura72}, \citet{shakura73}, and
\citet{novikov73} and since then it has been  widely used for
geometrically thin and optically thick accretion disks in  moderate
to high luminosity AGNs.  In this model the viscous transport
coefficient is limited by the vertical size of an eddy that can "fit"
within the height of the disk, $2H$, and the velocity of the eddy of
less than sound speed, $c_s$, within the disk. Thus the maximum
possible viscous transport coefficient, $\nu_{max}$ becomes 
$\nu_{max} < H c_s$, regardless of what source of turbulence or
instability one invokes.  The consequence of this limitation is that
using the Shakura--Sunyaev formalism,  a constant mass flow and the
physics of radiation transport, pressure, and surface emission one
obtains  a disk around a typical CMBH of $10^8 M_{\odot}$ that has
too great a mass thickness at too small a radius, $\sim 0.013 $ pc to
be consistent within several orders of magnitude with a generally
accepted picture of galaxy formation and angular momentum
distribution of a "flat rotation curve"  disk. This difficulty has
been recognized for some time,
\citep{shlosman89}, motivating the consideration of various 
alternate transport mechanisms.  However, a recent in-depth review
of the problem by
\citet{goodman03}  finds no simple solution.

   As an alternative solution we have found in recent years that
large scale horizontal vortices can be excited within a Keplerian
disk by appropriate pressure or angular momentum distributions,
closely analogous to Rossby vortices within the disk \citep{li01b}. 
These vortices initially have a horizontal dimension of
$\sim 2 \; \mbox{to}\; 4 \; H$. 
One might then ask what is the difference with the truncation of eddy
size at the disk height of a turbulent disk and the Rossby vortex
disk, because both are truncated initially at the same size. The
difference is that the Rossby vortices act coherently and so each
vortex, regardless of size acts to transport angular momentum in one
direction only, namely radially outwards as compared to turbulence,
which is a random walk process.  Furthermore the Rossby vortices
have a further property of merging leading to larger vortices until 
$r_{vortex} \simeq R/3$.
The transport process is then faster  or a transport coefficient that
can be larger by the ratio
$\nu_{Rossby} / \nu_{turbulence} \simeq r/H \sim 10^4$, thus making
feasible an accretion disk that matches the flat rotation curve mass
and angular momentum distribution  of typical galaxy formation.  In
addition we also take note of the fact that we  have recently
suggested that the origin of CMBHs and their correlated power  law
velocity dispersion can be surprisingly  explained by forming the
CMBH  accretion disk using the Rossby vortex  instability mechanism
rather than the  Shakura--Sunyaev turbulent  model
\citep{colgate03}. This prediction and confirmation by observations
as well as the mass thickness problem  is sufficiently provoking
that to consider the  accretion disk dynamo model based solely upon
the Shakura--Sunyaev model may  be misleading. Fortunately the 
Rossby vortex instability predicts  universally a thinner disk and
all disk problems with the dynamo  become less difficult. Still, as
it is described in a companion paper~II \citep{pariev06}, star-disk
collisions driven dynamo operates at radii
$\sim 200 r_g$ in the accretion disk, where too high mass thickness
of Shakura--Sunyaev disk is not yet a problem for self-gravity and
matching to outside "flat rotation curve". Shakura--Sunyaev model is
also better developed than Rossby vortex model at present. Hence, in
order to minimize the number of speculative assumptions, we proceed 
with our dynamo model based  upon the Shakura--Sunyaev disk model 
and note the alternate  differences  when necessary.

This  work is arranged as  follows: in section~\ref{sec2} we discuss
the distribution of stars, in section~\ref{sec3} the  structure of
the accretion disk, and in section~\ref{sec4} the kinematics of
star-disk collisions. Finally, we end with a summary.

\section{Star Clusters,  and their Distributions}
\label{sec2}

To proceed with the dynamo problem we need to address the following
issues:

\begin{enumerate}
\item What is the distribution of stars in coordinate and velocity
space  in the central star cluster of an AGN~?
\item What is the velocity, density  and conductivity of the plasma
in  the disk and in the corona of the disk~?
\item What is the hydrodynamics of the flow resulting from the
passage  of the star through the disk~?
\end{enumerate}

Each of these problems is difficult to solve. Moreover, there are no
detailed solutions to these problems up to date. Furthermore they
all interrelate. In the following three subsections we present a
brief (far from complete) analysis of each of the problems based on
available research and some of our own conjectures. Because each of
these problems  interrelate to some degree with each other, the
justification of some assumptions must be delayed.  However, as
noted above, we will predict a dynamo gain so large that details of
the disk and of the star disk collisions and their frequency  become
of secondary importance compared to the existence of the disk, a few
stars and the CMBH.

\subsection{Kinematics of the Central Star Cluster}
\label{subsec_kinem}

By now there is  strong observational evidence (e.g.,
\citealt{tremaine02};
\citealt{merritt01};
\citealt{vandermarel99};
\citealt{kormendy98};
\citealt{vandermarel97}) that many galactic  nuclei contain massive
dark objects in the range of
$\approx 10^6-10^9\,M_{\odot}$. Numerical simulations of the
evolution of central dense stellar clusters indicate that they are
unstable to the formation of black holes, which would subsequently
grow to larger masses by absorbing more stars  \citep{quinlan90}.
Recent observations and the interpretation of  very broad skewed
profiles of iron emission line (e.g.,
\citealt{tanaka95}; \citealt{bromley98};
\citealt{fabian00}) in Seyfert nuclei provide direct evidence for
strong gravitational effects in the vicinity of massive dark objects
in AGNs. This leaves us with conviction that the nuclei of AGNs
indeed harbor black holes with accretion disks \citep{fabian95}.
Although the observations of star velocities and velocity dispersion
are used to obtain an estimate of the  mass of the supermassive black
hole, a measurement of the number density of stars is limited by
resolution to about 1~pc for M32 and M31 and about 10~pc for  the
nearest ellipticals. From these observations we infer a star density
of
$n(\mbox{1 pc})\approx 10^4 - 10^6\, M_{\odot}
\mbox{pc}^{-3}$ at 1~pc \citep{lauer95}.

One needs to rely on the theory of the evolution of the central star
cluster  in order to obtain  number densities of stars closer  to the
black hole. The subject of the evolution of a star cluster around a
supermassive black hole has drawn significant interest in the past.
The gravitational potential inside of the  central 1~pc will be
always dominated  by the black hole. \citet{bahcall76} showed that,
if the evolution of  a star cluster is dominated by relaxation, the
effect of a central Newtonian point mass on an isotropic cluster
would be to create a density profile
$n \propto r^{-7/4}$. However, for small radii ($\approx
0.1-1\,\mbox{pc}$) the effects of physical collisions between stars
become dominant over two-body relaxation. Also, the  disk produces a
drag on the stellar orbits, which accumulates over many star passages.
The result of the star-disk interactions is to reduce the inclination,
eccentricity, and semimajor axis of an orbit, finally causing  the
star to be trapped in the disk plane, and so moving on  circular
Keplerian orbits
\citep{syer91, artymowicz93, artymowicz94, rauch95, vokrouhlicky98}.
Closer to the black hole ($\le 100 r_g
$, $r_g = 2GM/c^2$, the gravitational radius) general relativistic
corrections to the orbital motions and tidal disruption of the 
stars by the black hole must be taken into account. Considering all 
these effects and furthermore that the star-star collisions cannot be
treated in a Fokker--Plank (or diffusion) approximation, an accurate
theory becomes  a difficult endeavor, which has not yet been
completed  to our knowledge.

To obtain a plausible estimate of the number density and velocity
distribution of stars in the central cluster we will follow the
work of \citet{rauch99}, which addresses all these effects on
the star distribution mentioned above, except the dragging by the
disk. \citet{rauch99} showed that star-star collisions lead to the
formation of a plateau in the density of stars for small
$r$ because of the large rates of destruction of stars by collisions.
We adopt the results of model~4 from
\citet{rauch99} as our fiducial model. This model was calculated for
all stars having initially one solar mass. The collisional evolution
in model~4 are close to the stationary state, when the combined
losses of stars due to collisions, ejection, tidal disruptions and
capture by the black hole are balanced by the replenishment of stars
as a result of two-body relaxation in the outer region  with
$n\propto r^{-7/4}$ density profile. Taking into account the order
of  magnitude uncertainties in the observed star density at 1~pc, the
fact that model~4 has not quite reached a stationary state can be
acceptable for the purpose of order of magnitude estimates.

For the mass of the black hole we take
$M=10^8 M_8
\,M_{\odot}$. The radius of the event horizon of the black hole is
$r_g=2GM/c^2=3.0\cdot10^{13}\cdot M_8
\,\mbox{cm}= 9.5\cdot 10^{-6}\cdot M_8
\,\mbox{pc}$. We then approximate the density profile of model~4 as
\begin{eqnarray} && n=n_5 \cdot 10^5 \,
\frac{M_{\odot}}{\mbox{pc}^3}\,
\left(\frac{r}{1\mbox{pc}}
\right)^{-7/4} \quad \mbox{for} \quad r>10^{-2}\,\mbox{pc}\mbox{,}
\nonumber \\ && n=n_5 \cdot 3\cdot
10^8\,\frac{M_{\odot}}{\mbox{pc}^3}
\quad
\mbox{for}
\quad 10r_t<r<10^{-2}\,\mbox{pc}\label{eqn3.1}\mbox{,}
\\ && n=0 \quad \mbox{for} \quad r<10r_t\mbox{,}
\nonumber
\end{eqnarray} where $r_t=2.1\cdot 10^{-4}\,\mbox{pc}\cdot M_8^{1/3}
= 21 r_g$ is the tidal disruption radius for a solar mass star,
$\displaystyle n_5=\frac{n(1\,\mbox{pc})}
{10^5\,M_{\odot}/\mbox{pc}^{-3}}$,
$M_8=M/10^8 M_{\odot}$. An integration of expression~(\ref{eqn3.1})
over volume produces the number of stars with impact radii inside a
given radius,
$N(<r)$, as:
\begin{eqnarray} && N(<r)=n_5 \cdot\left[  10^6 \,
\left(\frac{r}{1\mbox{pc}}
\right)^{5/4}-1.9\cdot 10^3 \right]  \quad \mbox{stars for}
\quad r>10^{-2}\,\mbox{pc}\mbox{,}
\nonumber \\ && N(<r) = n_5 \cdot 12 \left[
\left(\frac{r}{10 r_t}\right)^{3} -1
\right]
\quad
\mbox{stars for}
\quad 10r_t<r<10^{-2}\,\mbox{pc}\label{eqn3.2}\mbox{,}
\\ && N(<r)= 0 \quad
\mbox{stars for}
\quad r< 10r_t\mbox{,}
\nonumber
\end{eqnarray} such that $N(<10^{-2}\,\mbox{pc})=1.3\cdot 10^3\,n_5$
stars. Thus in this conservative view there are no star disk
collisions and therefore no dynamo inside $200 r_{g}$. One notes
that the total mass of stars inside central
$0.1\,\mbox{pc}$ remains a small fraction ($< 10^{-2}$) of CMBH mass.

This extrapolated lack of stars within the inner most regions of the
disk presumably occurs because of star-star  collisions and tidal
disruption of stars and is independent of disk structure. The zero
$n$ at $r<10 r_t$ is a crude approximation to actual decrease in
the number density of stars. This is because we recognize that
distant gravitational scattering will lead to some diffusion of
stars from distant regions and thus feeding of stars to the inner
regions, limited by
$r_{t}$.

We shall comment further on the influence of the drag by the disk on
the  above density profile. Following the formula~[1] from
\citet{rauch99} the probability that the solar mass star on the
elliptic orbit  with eccentricity
$e$ and the minimum distance from the black hole
$r_{min}$ will experience a collision with  another star during one
orbital period is
\begin{equation}
\tau_{coll}=2\cdot 10^{-5}\cdot n_5
M_8^{-3/4}(3-e)\left(\frac{r_{min}}{r_g}
\right)^{-3/4}\label{eqn3.3}\mbox{.}
\end{equation} This probability at $100 r_g$ or
$\sim 10^{-3}\,\mbox{pc}$ becomes
\begin{equation}
\tau_{coll}=6\cdot 10^{-7}\cdot n_5
M_8^{-3/4}(3-e)\label{eqn3.4}\mbox{.}
\end{equation} This probability is sufficiently small that the  drag
of the disk during star-disk collisions can be more important. In
order to evaluate that drag we need to know the surface density in the
disk.

\section{Disk Structure and Star Collisions}
\label{sec3}

We adopt the $\alpha$-disk model, which we also refer to as the
Shakura--Sunyaev 
\citep{shakura72, shakura73} model.  We also consider the Rossby
vortex model for reasons outlined in the introduction. As noted
before, fortunately the  Rossby vortex instability predicts 
universally a thinner disk and all disk problems with the dynamo
become less difficult.  Hence we proceed  with our dynamo model
based  upon the Shakura--Sunyaev disk model  and note the alternate 
differences  when necessary. 

For thirty years, the Shakura--Sunyaev disk model has been the most
widely used model of the accretion disk. The expressions for the
parameters of the $\alpha$-disk can be found in original articles
\citep{shakura72, shakura73} and in many later books (e.g.,
\citealt{shapiro83, krolik99, bisnovatyi02}). Here, we give the
complete set of these expressions conveniently scaled for our 
problem (supermassive black hole, radius about $200 r_g$  or
$10^{-2}\,\mbox{pc}$) in Appendix~A.

There have been a number of works
perfecting and improving the simple analytical  Shakura--Sunyaev
model and determining the limits of applicability of this solution to
real AGN accretion disks.  Here we leave aside the complex physics of
the innermost ($\leq 10 r_g$) parts of  the accretion flow because
the innermost regions are devoid of stars and so star-disk
collisions are almost non-existent in this region.  More realistic
bound-free opacities were included by
\citet{wandel88}, non-LTE models were developed by \citet{hubeny97,
hubeny98} in disks with arbitrary optical depth, and optically thin
and optically thick disks, were considered in \citet{artemova96}. If
one is looking at the interval of disk radii $\sim 100$ to $\sim 1000
r_g$, these improvements have some quantitative effects on the disk
structure such as the emitted spectrum may be significantly
different among models.  More exact descriptions of the accretion
disk come at a price of loosing analytic  simplicity of the
expressions for the radial profiles of the density, temperature, disk
height, etc., while gaining a factor of only a few in
accuracy. Because of the approximate  nature of our model (mandated
by the poor accuracy of its other ingredients), we prefer to use the
simplest of the disk models, and therefore use the analytic results
given in the original works of  Shakura and Sunyaev.

The surface  density of the
$\alpha$-disk in the inner radiation dominated part, where Compton
opacity prevails, is given by expression~(\ref{Sigma_a}) in the
Appendix~A. When expressed in units of $M_{\odot}/R_{\odot}^2$, it
becomes
\begin{equation}
\Sigma = 9.9\cdot 10^{-10}\,\frac{M_{\odot}}{R_{\odot}^2}
\left(\frac{\alpha_{ss}}{0.01}\right)^{-1}
\left(\frac{l_{E}}{0.1}\right)^{-1}\left
(\frac{\epsilon}{0.1}\right)^{1}\left
(\frac{rc^2}{GM}\right)^{3/2}\left (1-\sqrt{\frac{3r_g}{r}}
\right)^{-1}
\label {eqn3.5}\mbox{,}
\end{equation} where $\alpha_{ss}$ is the ``$\alpha$''-parameter of
the disk model,
$l_E$ is  the ratio of the luminosity of the disk to the Eddington
limit for the black hole of mass $M$,
$\epsilon$ is the fraction of the rest mass energy of the accreting
matter, which is radiated away. Thus close to $r_g$, $\Sigma
=404\,\mbox{g cm}^{-2}$. The expression~(\ref{eqn3.5}) is valid for a
radiation pressure supported disk where $r<r_{ab}$ given by
expression~(\ref{rab}). For typical values
$\alpha_{ss}=0.01$,
$\epsilon=0.1$, $l_E=0.1$, $M_8=1$,  we obtain
$r_{ab}=2.3\cdot 10^{-3}\,\mbox{pc}\approx 240 r_g$ and
$\Sigma_{ab} = \Sigma(r_{ab}) \approx 4.2 \cdot 10^6 \,\mbox{g
cm}^{-2}$.

When the disk becomes self gravitating, it may become subject to a
gravitational instability. In Appendix~A we check that by calculating
the Toomre parameter $\displaystyle \mbox{To}=\frac{\varkappa
c_s}{\pi G \Sigma}$ (e.g., \citealt{binney94}), where $\varkappa$ is
the epicyclic frequency and $c_s$ is vertically averaged sound speed.
The gravitational instability develops if $\mbox{To}<1$. As follows
from the analysis in the Appendix~A the disk has a well defined
radius of stability $r_T$, such that for $r>r_T$ it becomes unstable.
In the case when $r_T<r_{ab}$, the expression for $r_T$ is given by
formula~(\ref{r_Q}). For the values $\alpha_{ss}=0.01$,
$\epsilon=0.1$,
$l_E=0.1$, $M_8=1$ the radius of stability $r_T$ falls close to the
radius of transition $r_{ab}$ between radiation dominated and gas
pressure dominated parts of the disk. The development of the Jeans
instability should lead to the formation of spiral patterns and
fragmentation of the disk
\citep{shlosman89}, which will happen on the radial inflow time
scale at a radius
$\approx r_T$. Therefore, for estimating the drag produced by the
disk on the passing stars, we can limit ourselves to consider only
the inner portion of the disk at $r<r_{ab}$ and use
equation~(\ref{eqn3.5}) for the disk surface density. The gas beyond
$r_{ab}$ may also influence the motion of stars. It is difficult to
evaluate the drag produced on stars passing through gravitationally
unstable outer parts of the disk for $r>r_T$. However, we note that
the rate of star-disk collisions is maximized at $r\la 10 r_t \sim
r_{ab}$, so most of the star-disk collisions happen inside the
radiation dominated zone (zone~(a)) of the disk.

The Rossby vortex model of the disk
    predicts a mass thickness of a near constant,
$100\,\mbox {g cm}^{-2} <\Sigma_{RVI} < 1000\,\mbox{g cm}^{-2}$.
This is about the same as the Shakura--Sunyaev model near to the BH,
but becomes very much less at large radius.  Consequently  the self
gravity condition occurs at a much larger radius, $3$ to
$10\,\mbox{pc}$, and matches smoothly  onto the galactic flat
rotation curve mass  distribution.

Hereafter, we will use disk parameters in zone~(a) listed in
Appendix~A for the estimates of star-disk collisions. The disk
half-thickness (expression~(\ref{H_a})) expressed in units of
$r_g$ is
\begin{equation} H = 1.15 \cdot  r_g \left(\frac{l_E}{0.1}
\right)
\left(\frac{\epsilon}{0.1}
\right)^{-1}M_8\left(1-\sqrt{\frac{3r_g}{r}}
\right)\label{H_rg}\mbox{,}
\end{equation} expressed in solar radii
\begin{equation} H=370 R_{\odot}
\left(\frac{l_E}{0.1}\right)\left(
\frac{\epsilon}{0.1}\right)^{-1} M_8
\left(1-\sqrt{\frac{3r_g}{r}}
\right)\label{H_solar}\mbox{,}
\end{equation} and expressed as a fraction of
$r_{ab}$
\begin{equation} H=3.7\cdot 10^{-3} r_{ab}
\left(\frac{\alpha_{ss}}{0.01}\right)^{-2/21}
\left(\frac{l_E}{0.1}\right)^{5/21}
\left(\frac{\epsilon}{0.1}\right)^{-5/21} M_8^{-2/21}
\left(1-\sqrt{\frac{3r_g}{r}}\right)
\label{H_rab}\mbox{.}
\end{equation}

It is natural to expect that the dynamo growth rate will be also
maximized at small radii primarily within zone (a) where the disk is
radiation dominated, but outside of the region, $r_t
\simeq 21 r_g$, of tidal destruction of stars. However, we  should
also point out that although proof of principle of the dynamo is
most likely where the growth rate is maximum, we also expect that
regardless of where the growth rate maximizes,  that  the back
reaction  will limit the maximum fields  and that subsequent
diffusion outwards (as for the angular momentum)  and advection
inwards (as for the mass)  will ensure a redistribution of the
magnetic flux reaching a new equilibrium presumably less dependent
upon  where the maximum dynamo growth rate  occurs.

\subsection{Star-Disk Interaction}
\label{subsec_3.3}

The orbital period of the star is
\begin{equation} t_{orb}=3.1\cdot 10^3\,\mbox{s}\,\cdot M_8
\left(
\frac{r_{min} c^2}{GM}\right)^{3/2}
(1-e)^{-3/2}\label{eqn3.9}\mbox{.}
\end{equation} where, as before,
$r_{min}$ is the minimum impact radius of the star's orbit.  The
typical velocity of the star relative to the disk is close to the
Keplerian velocity at
$r_{min}$. Since the speed of sound in the  disk is much smaller
than the Keplerian velocity, by the ratio $H/r \simeq 3.7 \cdot
10^{-3}$, stars pass through the disk with highly supersonic
velocities. The drag force on the star consists of two components,
collisional and gravitational. The collisional or direct drag is
produced by intercepting the  disk material by the geometric cross
section of the star. Assuming the star to have a solar mass and
radius, this force is
$F_{drag}=\pi R_{\odot}^2\rho v_{\ast}^2$,  where
$\rho$ is the mass density of the gas in the disk, and
$v_{\ast}$ is the velocity of the star relative to the disk gas.
Radiation drag is negligible compared to gas drag as soon as  the
speed of sound is nonrelativistic, i.e.
$c_s\ll c$. The second component of the drag force is due to
deflection of the gas by the gravitational field of the star.
\citet{rephaeli80} found that the latter component  is nonzero only
for supersonic motion and gave the following expression for that
force in the limit
$v_{\ast}\gg c_s$
\begin{equation} F_{grav}=4\pi
\frac{G^2 M_{\odot}^2}{v_{\ast}^2}\rho
\ln\Lambda
\label{eqn3.10}\mbox{,}
\end{equation} where $\Lambda$ is the Coulomb logarithm. The ratio
of the two forces is
\begin{equation}
\frac{F_{drag}}{F_{grav}}=\frac{R_{\odot}^2 v_{\ast}^4}{G^2
M_{\odot}^2\, 4
\ln\Lambda}\label{eqn3.11}\mbox{.}
\end{equation} Using for
$v_{\ast}$ its Keplerian value
$v_{\ast}=(GM/r)^{-1/2}$, and using for the Coulomb logarithm its
maximum possible value
$\Lambda=r/R_{\odot}$, one obtains the ratio of the  forces as
\begin{equation}
\frac{F_{drag}}{F_{grav}}=\frac{1.03\cdot 10^{10}}{1+0.19\ln(M_8
\frac{c^2 r}{GM})}
\left(\frac{GM}{c^2r}\right)^2
\label{eqn3.12}
\mbox{.}
\end{equation}

One can see from equation~(\ref{eqn3.12}) that the force due to the 
direct interception of gas by the star is much larger than the drag 
caused by the gravitational drag for all values of
$r$ of interest  to us $r\la 10^5 r_g$. Thus, we can consider the
change of momentum  caused by the disk on passing stars as purely due
to the interception of  the gas by the geometrical cross section of
the star
$\pi R_{\odot}^2$. Hence, the characteristic time needed to
substantially change the star orbit as a result of star-disk
interactions,
$t_{disk}$, is approximately equal to the time needed for the star to
intercept the disk mass equal to the mass of the star. A star will
pass through the disk twice per one  orbital period. Assuming all
stars as having a solar mass and radius, the ratio of the orbital
period to
$t_{disk}$ is
$$
\tau_{disk}=\frac{t_{orb}}{t_{disk}}\approx
\frac{2\Sigma \pi R_{\odot}^2} {M_{\odot}}\mbox{.}
$$ Using expression~(\ref{eqn3.5}) for
$\Sigma$ in the region $r<r_{ab}$  one obtains
\begin{equation}
\tau_{disk}=6.2\cdot 10^{-9}\,
\left(\frac{\alpha_{ss}}{0.01}\right)^{-1}
\left(\frac{l_E}{0.1}\right)^{-1}\frac{
\epsilon}{0.1}\left(\frac{c^2 r_{min}}
{GM}\right)^{3/2}\left(1-\sqrt{\frac{3r_g}
{r_{min}}}\right)^{-1}\label{eqn3.13}
\mbox{.}
\end{equation} The corresponding star-disk interaction time scale
$t_{disk}$ is given by
\begin{equation} t_{disk}=1.58\cdot 10^4
\,\mbox{yr}
\cdot
\frac{\alpha_{ss}}{0.01}\frac{l_E}{0.1}
\left(\frac{\epsilon}{0.1}\right)^{-1} M_8^{-1/2}
\frac{1}{(1-e)^{3/2}}
\left(1-\sqrt{\frac{3r_g}{r_{min}}}\right)
\label{eqn3.14}\mbox{,}
\end{equation} and is independent of the semi-major axis of the star
orbit. As was shown by \citet{rauch95} secular evolution of all
orbital elements of a star happen at the same time scale
$t_{disk}$ from equation (\ref{eqn3.14}). The ratio of
$\tau_{disk}$ to
$\tau_{coll}$ (equation~(\ref{eqn3.3}))  is given by
\begin{equation}
\frac{\tau_{disk}}{\tau_{coll}}= 1.8\cdot 10^{-4}\,
n_5^{-1}\frac{1}{3-e}
M_8^{3/4}\left(\frac{\alpha_{ss}}{0.01}\right)^{-1}
\left(\frac{l_E}{0.1}\right)^{-1}\left(
\frac{\epsilon}{0.1}\right)^{1}
\left(\frac{c^2 r}{GM}\right)^{9/4}\left(1-\sqrt
{\frac{3r_g}{r_{min}}}\right)^{-1}
\label{eqn3.15}\mbox{.}
\end{equation} For orbits with
$r_{min}\le 30 r_g$ one has
$\tau_{disk}< \tau_{coll}$ and  the effect of star-star collisions
dominates over the effect of star-disk collisions (assuming typical
parameters for the disk). For the radii $30 r_g \le r_{min} \le r_T$
the orbit evolution is  more influenced by the drag from the disk
rather than by star-star collisions.  (We note that this radius,
$30r_g$,  is only slightly greater than the gravitational disruption
radius by the CMBH,
$r_{t}
\simeq 20 r_g$.) Only a fraction of stars from the outer region
located beyond
$\approx 1000 r_g$ will not be put into the disk plane by star-disk
drag. Results of \citet{rauch95} show that it takes a considerably
longer  time than
$t_{disk}$ to reorient the retrograde star orbits. During this 
reorientation process the semimajor axes of initially retrograde
star orbits decreases by
$\approx 10$ times. Before the alignment process for such stars
could be completed they will move in radius closer than
$\approx 30 r_g$ into the star-star collisions zone, where their
orbital inclinations would be randomized. Another factor preventing
all stars from being trapped into the disk plane is that there is
always a fraction of stars which  are injected by two body relaxation
into the neighborhood of the black hole from large (much larger than
$r_T$) radii. These stars can be brought directly into the region $r
\le 30 r_g$ (or close to it) and contribute to the collisional core
of the stellar cluster.

To summarize, both star-disk and star-star collisions can be
important for  determining the distribution function in the central
star cluster. However, it seems unlikely that the drag by the disk
can trap all stars into the disk plane and denude the central $\approx
10^{-3}\,\mbox{pc}$ of all stars not in the disk plane. Trapping of
stars by the disk will reduce the numbers of stars given by
(\ref{eqn3.1}) but this requires more evolved computations, which
are beyond the scope of the present work.  Both star-star collisions
and the effect of trapping by the disk  of the stars having lower
eccentricities faster than the stars having  larger eccentricities
leads to highly eccentric orbits of stars in the central
$\approx 10^{-3}\,\mbox{pc}$. Drag by the disk will also lead to  the
prevailing of  prograde orbits over the retrograde orbits.  However,
for our purpose, we assume that the star density is given by
equations~(\ref{eqn3.1}), all stars have
$e=1$ and their orbits are randomly oriented in space. (This
approximation is better in the model of the disk driven by Rossby
vortices.)

\subsection{The Rate of Star-Disk Collisions}
\label{subsec_stdiskrate}

We shall use the  number density of stars, $n$, given by
equation~(\ref{eqn3.1}) in order to evaluate the rate of star-disk
collisions. The flux of stars through the disk coming from one side
of it is
$nv/4$, where we assume that all stars have the same velocity
$v=\sqrt{2} (r\Omega_K)$  (parabolic velocity) and are distributed
isotropically. One obtains then for
$M_8=1$
\begin{eqnarray} && \frac{1}{4}nv=2.4
\cdot 10^{-39}\,\frac{1}{\mbox{cm}^2\mbox{s}}\,
n_5\left(\frac{r}{10^{-2}\,\mbox{pc}}
\right)^{-9/4}
\quad
\mbox{for}
\quad r>10^{-2}\,\mbox{pc}\mbox{,}
\nonumber \\ &&
\frac{1}{4}nv=2.4
\cdot 10^{-39}\,\frac{1}{\mbox{cm}^2\mbox{s}}\,
n_5\left(\frac{r}{10^{-2}\,\mbox{pc}}
\right)^{-1/2}
\;
\mbox{for}
\; 10r_t<r<10^{-2}\,\mbox{pc}
\label{sc_eqn3.16}\mbox{,}
\\ &&
\frac{1}{4}nv=0 \quad \mbox{for}
\quad r<10r_t\mbox{.}\nonumber
\end{eqnarray} Integrating the flux of stars coming from  both sides
of the disk over an area of $\pi r^2$ inside some given  radius $r$,
one can estimate the rate of  star-disk collisions within the radius 
$r$. Let us define the time $\Delta T_c=\Delta T_c(r)$ as the
inverse  of this rate, i.e. one star passes through the disk area
inside the radius $r$ during the time
$\Delta T_c$ on average. The result is (see equation~(\ref{eqn3.1}))
\begin{eqnarray} && \Delta T_c=\frac{2\pi}{\Omega_K (r)}\cdot
2.8\cdot 10^{-5}\cdot
n_5^{-1}\left(\frac{r}{10^{-2}\,\mbox{pc}}\right)^{-3/2}
\quad
\mbox{for} \quad r>10^{-2}\,\mbox{pc}\mbox{,} \nonumber \\ &&
\Delta T_c=\frac{2\pi}{\Omega_K(r)}\frac{1.9\cdot 10^{-2}}{n_5
\left(\frac{r}{10 r_t}\right)^{3/2}\left(\left(\frac{r}{10 r_t}
\right)^{3/2}-1\right)}
\; \mbox{for} \; 10r_t<r<10^{-2}\mbox{pc}\label{sc_eqn3.17}\mbox{,}
\\ && \Delta T_c=\infty \quad \mbox{for} \quad r<10r_t
\quad
\mbox{(no collisions)}\mbox{,}
\nonumber
\end{eqnarray} where
$2\pi/\Omega_K(r)=T_K(r)$ is the period of Keplerian circular orbit
at the radial distance $r$ from the black hole. We see that the
number of star-disk collisions happening per Keplerian period,
$T_K(r)$, is $\propto r^3$ inside the collisional core of the star
cluster, e.g. within
$\approx 10^{-2}\,\mbox{pc}$. For the outer region of the stellar 
cluster beyond
$\approx 10^{-2}\,\mbox{pc}$ this number continues to increase with
$r$ but more slowly, as $\propto r^{3/2}$. The number of collisions
per Keplerian period at 0.01 pc is $\sim 30,000$,  leading to
fluctuations of the order of $1\%$ within an orbital time of several
years.

If these collisions should produce broad emission and absorption
lines regions, (BLRs), then this result may not be inconsistent with
observations.  Estimates of the density of the matter leading to the
broad emission lines from the interpretation of allowed and forbidden
transitions give a  density of $\rho_{BL} \sim 10^{-11}\> \mbox
{to}\> 10^{-13}\,
\mbox{g}/\mbox{cm}^3$, \citep{sulentic00}.   The geometrical
thickness of the disk 
$H$ in radiative pressure dominated inner zone is independent of the
disk model and the mechanism of angular momentum  transport and is
given by equation~(\ref{H_a}). In thermal pressure dominated part of
the disk, $H$ weakly depends on $\Sigma$  as $H\propto
\Sigma^{1/8}$.  Only in the case of the RVI disk does the low
thickness, $\Sigma_{RVI} \sim 10^2 \> \mbox {to}\> 10^3 \,
\mbox{g}/\mbox{cm}^3 $, lead to a sufficiently low density,
$\rho_{RVI} = \Sigma_{RVI}/H \simeq
\Sigma_{RVI} \cdot 3 \cdot 10^{-14} \, \mbox{g}/\mbox{cm}^3$, which
is consistent with the above estimates for the density of the
star-disk driven matter emitting the broad emission lines. On the
other hand, the Shakura--Sunyaev disk would be expected to have a
density
$\rho_{SS}$ given by expression~(\ref{rho}) in the radiation 
dominated zone~(a) and expression~(\ref{rho_zoneb}) in the  pressure
dominated zone~(b). If one equates the observed width of broad
emission lines ($\sim 7\cdot 10^3 \,\mbox{km/s}$) to the Doppler
shift at Keplerian velocity, one obtains an estimate of the location
of the broad lines region at $r\sim 10^3 r_g$. This radius falls not
far from the boundary between zones (a) and (b) in the 
Shakura--Sunyaev disk model (see expression~(\ref{rab})). The
density of the  Shakura--Sunyaev disk at this radius is $\sim
10^{-6}\,
\mbox{g}/\mbox{cm}^3$  to $10^{-8}\, \mbox{g}/\mbox{cm}^3$ depending
upon the parameters of the model.  This is at least $5$ orders of
magnitude larger than $\rho_{BL}$  required by observations.  The
differences in $\rho$ for Shakura--Sunyaev  and RVI disks are almost
completely attributable to the much  lower column thickness
$\Sigma_{RVI}$ than $\Sigma$  for Shakura--Sunyaev model.
Regardless, the function of the plumes for producing the helicity for
the dynamo should be independent of these differences in the models
of the disk.

Star disk collisions  were first suggested as the source of the
BLRs  by
\citet{zurek94}, \citet{zurek96}, but a detailed calculation of the
phenomena has not yet been performed, because it requires 3-D
hydrodynamics with radiation flow and opacities determined by
multiple lines. An approximation to this problem was calculated by
\citet{arm96} for the purpose of determining the mass accretion rate
of giant stars by dynamic friction with the disk, but the radiation
flow in thin disks was not considered.  We recognize that very many
additional variables of hydrodynamics, radiation, and geometry must
be taken into account in order to positively identify BLRs with star
disk collisions.   With these caveats we proceed to analyze the star
collisions with the disk and  the resulting  plume formation  from
the standpoint of the fluid dynamics that has consequences for the
dynamo. 

\section{Plumes Produced by Star Passages through the Disk}
\label{sec4}

The first result of  a star-disk collision is to cause a local
fraction of the mass of the  disk to rise above the surface of the
disk because of the heat  generated by the collision. Two plumes
expanding on both sides of the accretion disk will be formed. A
second result is the expansion  of this rising  mass  fraction
relative to its vertical axis in the  relative vacuum above the disk
surface and again because of the  internal heat generated by the
collision. A third result is the  rotation (anticyclonic) of this
expanding matter relative to the  Keplerian frame corotating with the disk
because of the Coriolis force acting on  the expanding matter.  Again
we emphasize that this rotation through a finite angle has been
measured in the laboratory and agrees with a simple theory of
conservation of angular momentum and radial expansion of the plume
\citep{beckley03}.  All three effects are important to the dynamo 
gain.  However, we will find that the dynamo gain during the life 
time of the accretion disk, $\sim 10^8$ years, is so large that the
accuracy of the detailed description of these "plumes" becomes of 
less importance compared to the facts of: (1) their axial
displacement  well above the disk; (2) their finite, $\sim \pi/2$
radians,  coherent rotation  every star-disk collision; and (3)
their  subsidence back to the disk in
$\sim \pi$ radians. In this spirit we  will estimate the
hydrodynamics of the star-disk  collision, attempting to establish
the universality of this phenomena  as the basis of the accretion
disk dynamo. As far as we know no hydrodynamic simulations of the
behavior of the disk matter due to  stars passing through the disk
have  yet been performed. (This is  because of the difficulty of
3-dimensional hydrodynamics with  radiation flow.) The star passes
through the disk at a  velocity, close to the Keplerian velocity of
the disk at  whatever radius the collision happens. The sound speed
in the  accretion disk is much  less than the Keplerian speed $v_K$:
$c_s
\simeq v_K H/r \simeq 3 \cdot 10^{-3} \, v_K$ at
$r_{ab}$, where
$H$ is the disk half-thickness given by expression~(\ref{H_a}) in 
zone~(a). Hence, the star-disk collisions are highly supersonic.
The  temperature of the gas in the disk, shocked by the star moving
at a  Keplerian velocity, is of the order of the virial temperature
in the gravitational potential of the central black hole. This
pressure must include the radiation contribution, which in general,
will be much larger than the particle pressure.  Because of the high
Mach number of the collision, the pressure of the shocked gas is very
much greater than the ambient pressure in the  disk. This over
pressure will cause a strong, primarily  radial shock, radial from
the axis of the trajectory,  in the wake of the star, because of the
large, length to diameter ratio of the hot channel, $H/R_{\odot}
\simeq 4\cdot 10^2$. After the star emerges above the disk surface
(i.e. higher than the  half thickness of the disk), the heated
shocked gas in the wake of the  star continues to expand sideways and
furthermore starts to expand vertically because of the rapidly
decreasing ambient pressure away from the disk mid-plane where the
pressure of the disk drops as
$\propto\exp(-z^2/H^2)$. Thus this expansion  can be treated as an
adiabatic expansion into vacuum after the plume rises by  a few
heights $H$ above and below the disk, provided the radiative loss is
fractionally small. We would now like to estimate the size, or
radius, $r_p$, of the matter that rises  $\sim 2H$ above the disk, or
to a height $l\simeq 3H$ above the mid-plane.  Although smaller mass
fractions with greater internal energy corresponding to smaller
radii of the shock will expand to greater heights above the disk,
nevertheless we are concerned with only this modest height, because
we expect that the mass and hence entrained magnetic flux to be
positively correlated with plume mass, and we wish to maximize the
entrained flux. On the other hand, by conservation of energy,  a
larger mass will rise or expand to a smaller height. We also desire
the plume to rise sufficiently above the disk such that there is 
ample time for radial expansion and hence torquing of the entrained 
magnetic field during the rise and fall of the plume material. This
will be our standard plume.

The radial  extend of the plume should be somewhat less than its
vertical extend because  the density gradient in the disk is largest
in the vertical direction. The  action of the Coriolis force leads
to  an elliptical shape of the horizontal cross section of the plume.
This is due  to the fact that epicycles of particles  in the
gravitational field of a point mass are ellipses with an axis ratio
of 2 and with an epicyclic frequency of
$\Omega_K$. We performed  simple ballistic calculations of
trajectories of  particles launched from a point at the mid-plane of
the disk with  initial velocities in different directions in the
horizontal plane. We obtained that at the time of maximum height of
the plume, $\approx T_K/4$, the position angle of the  major axis of
the ellipse is approximately
$-\pi/4$ from the outward radial direction
${\bf e}_r$. At the time of the fall back to the disk plane at
$\approx T_K/2$, the  major axis of the ellipse is close to the
azimuthal direction. Such a distortion in the shape  of an otherwise
cylindrical plume will only slightly  affect the rotation of the
entrapped toroidal flux and hence will not alter the dynamo action.

Before calculating the size or radius,
$r_p$, we first verify the adiabatic approximation  in that the
diffusion of radiation is fractionally small compared to the
hydrodynamic displacements. In this circumstance of a
Shakura--Sunyaev disk, this will allow us to treat the star-disk
collisions as strong shocks within the disk matter. Subsequently we
will consider the thinner, lower density Rossby disks
\citep{li01b} where radiation transport will dominate over shock
hydrodynamics.   However, for the purposes of the dynamo, the
production of helicity from either plumes will be similar.

\subsection{Radiation Diffusion in the Collision Shock}
\label{subsec_3.6}

During star-disk collisions the total energy taken from the star is
$\approx\Sigma v_K^2
\pi R_{\odot}^2$. This energy is distributed over a column of radial
extent,
$ \Delta R_{rad}$, due to radiation transport. For an estimate of
$\Delta R_{rad}$ one can take the distance from the star track where 
the sideways diffusion of radiation becomes comparable with the
advection of the radiation by the displacement of the disk matter
with the star velocity
$v_K$ (since the velocity of strong shock is of the order of
$v_K$). This results in
\begin{equation}
\Delta R_{rad}=\frac{c/3}{\kappa
\rho v_K}\mbox{,}
\label{eqn3.22}
\end{equation} where for $\rho$ we consider the density ahead of the
shock in the undisturbed disk matter to compare the radiation flux
with the transport of energy and momentum by the shock. We assume
$\kappa=0.4\,\mbox{cm}^2\mbox{g}^{-1}$, Thompson opacity. Then using
$\rho$ from expression~(\ref{rho}) at
$r_{ab}$,
\begin{equation}
\Delta R_{rad}= 10^{8}\,\mbox{cm}=1.4\cdot 10^{-3}
R_{\odot}\label{eqn3.23}
\mbox{.}
\end{equation} Since
$\Delta R_{rad}\ll R_{\odot}$, the radiation will remain local to the
shocked fluid. The state conditions in this shocked matter will
depend upon the rapid thermalization between the matter and
radiation.

The number of photon scatterings,
$n_{h\nu}$ within the time of traversal of the radiation front,
$\Delta R_{rad}$, becomes
\begin{equation} n_{h\nu} =  \left(
\Delta R_{rad}\kappa\rho
\right)^{2} = \left(
\frac{c/3}{v_K} \right)^{2}= 50
\left( \frac{r}{r_{ab}}\right)
\label{eqn3.24}
\mbox{.}
\end{equation} Therefore the radiation will be fully absorbed and 
thermalized with the  gas within $\Delta R_{rad}$. Since the gas
pressure is radiation dominated for $r < r_{ab}$ and the shock has a
high Mach number, $c_s/v_K \approx r/H
\simeq 280$ at
$r_{ab}$, then the shocked matter will have a still higher entropy
and be even further radiation dominated. In a strong shock  the
energy behind the shock will be half kinetic and half internal
energy, where in this case the radiation   pressure dominates. Thus
the subsequent evolution of the radiation dominated gas will be
governed by adiabatic hydrodynamics of the fluid with a polytropic
index $\gamma=4/3$.

\subsection {The Shock Produced by the Collision and Its Radial
Expansion}
\label{subsec_3.7}

Since the initial radius of the shocked gas is that of the star and
since this radius is small compared to the path length through the
disk, $H$, or
$H/R_{\odot}\approx 370$ at $r=r_{ab}$, we make the assumption that
the collision can be approximated as a line source of energy with the
energy deposition per unit length
$\Sigma v_K^2 \pi R_{\odot}^2$, and consider the  shock wave as
expanding radially from the  trajectory axis. This can be well
described as one of the sequence  of Sedov solutions
\citep{sedov59} of an expanding cylindrical shock in a uniform
medium. However, for the purposes of the accuracy required for our
plume approximation  it is sufficient to note that the energy
density left behind the shock,
$\epsilon_{shk}$, is nearly inversely proportional to the swept-up
mass, or
$\epsilon_{shk} \simeq
\epsilon_{shk,R\odot} (R_{\odot}/R_{shk})^2$ where
$\epsilon_{shk,R\odot} \simeq v_K^2/2$. This increase in energy
density leads to an increase in the pressure of the shocked gas
$P_{shk}$ relative to the ambient pressure $P_o$:
$P_{shk,R_{\odot}}(z) \simeq \rho v_K^2
\gg P_{o}(z)$ for all z. The high pressure of the shocked gas near
the axis of the channel will drive the shock to larger radii while
expanding adiabatically behind the shock.  Near the surface,
$R_{shk}
\simeq z$ the shocked gas can expand vertically as well as
horizontally. However, to the extent that when the shock is strong,
$R_{shk} \ll H$, the  radial shock will have decreased in strength
before the star reaches the  surface and the over pressure becomes
too small except for  a small mass fraction of the surface mass,
$\Delta z \simeq R_{shk}
\ll H$, that will  expand vertically above the disk surface.

However, a larger mass will expand above the disk due to buoyancy.
In this case the vertical momentum is derived primarily from the
difference of gravitational forces on the buoyant matter versus the
ambient matter. The buoyant force is proportional to the entropy
ratio. A strong shock leaves behind matter whose entropy is higher
than the ambient medium. Since the entropy change due to a shock wave
is third order in the shock strength
\citep{courant48, zeldovich67}, only strong shocks result in
significant changes in entropy.   In this limit the entropy change
$\Delta S$ from the ambient entropy $S_o$ is  $\Delta S / S_{o}
\propto
\Delta (P/\rho)/(P_o/\rho)
\simeq (P_{shk}/P_{o}) ((\gamma -1)/(\gamma + 1))$ where $\gamma$ is
the usual ratio of specific heats, and $\rho$ is the ambient density.
The compression ratio is
$\eta_{CR} = \rho_{shk}/\rho = (\gamma + 1)/(\gamma -1) = 7$ across
a strong  shock for $\gamma = 4/3$. Thus, for example, for a plume to
rise well above the disk requires  an estimated
$\Delta  S /S_{o} \geq 2 $ and thus
$P_{shk}/ P_{o} \simeq 14$ . Once the hot shocked gas rises to the
surface of the disk and assuming that this flow is adiabatic
thereafter and thus does not entrain a significant fraction of
surrounding matter, the subsequent expansion above the disk  is
determined by its initial internal energy.

Let us consider the neighborhood of a point
$r=r_0$ at the mid-plane of the  disk where a star disk collision
occurs. One can introduce a local Cartesian coordinate system
$x$, $y$, $z$ in the Keplerian rotating frame  with the origin at the
point
$r=r_0$ such that the $x$-axis is directed radially outward, the
$y$-axis is directed in the positive azimuthal direction, and the
$z$-axis  is perpendicular to the disk plane. Then, the effective
gravitational and  centrifugal potential in the Keplerian rotating
frame in the  neighborhood of the point $r=r_0$ is
\begin{equation}
\Delta\Phi=\frac{GM}{2r_0^3}(z^2-3x^2)\mbox{.}
\label{eqn3.21}
\end{equation} The thermal energy  of the hot column of gas is a
fraction of the loss of kinetic energy of the star due to the
hydrodynamic collision with the disk. This latter energy loss during
one passage is
$F_{drag}2H=2H\pi R_{\odot}^2\rho v_{\ast}^2=\pi R_{\odot}^2
\Sigma v_{\ast}^2$.

Without a hydrodynamic simulation in 3-dimensions an accurate
description is missing. Nevertheless it is sufficient to approximate
the solution as that fraction of the disk matter that has an internal
energy density,
$\epsilon_{shk}$ greater than that of the ambient disk by that
factor such that it will rise to a height, $z$, determined by its 
potential energy, or
$\displaystyle\Delta\Phi = \frac{GM z^2}{2r^3}$
(equation~(\ref{eqn3.21})). Since
$\Delta S/S_{o} =
\epsilon_{shk}/\epsilon_{o}$, where $\displaystyle\epsilon_{o}
\simeq \Delta\Phi(H) =
\frac{GM H^2}{2r^3}$,  then in order for a plume to rise above the
disk mid-plane to a height,
$l$,
\begin{equation}
\left(\frac {l}{H}
\right)^2
\simeq
\left(
\frac {\epsilon_{shk}}{\epsilon_{o}}
\right)
\simeq \left( \frac {v_K}{c_s}
\right)^2
\left( \frac{R_{\odot}}{R_{shk}}
\right)^{2}
\simeq  \left( \frac {r}{H}
\right)^2
\left(
\frac{R_{\odot}}{R_{shk}}
\right)^{2}
\label{eqn3.25}
\mbox{.}
\end{equation} We are concerned with plumes that rise well above the
disk so that they can expand horizontally by a factor several times
the plume's original radius. In this case  the moment of inertia of
the plume about its own axis will be increased by several times 
before falling back to the disk. This causes the plume to reduce its
own rotation rate relative to the frame of the disk, that is to
untwist relative to that frame. For this expansion to take place, the
plume must rise  roughly
$\sim 2H$ above the disk, or
$l\simeq 3 H$. At this height the pressure of the hydrostatic
isothermal atmosphere with the density profile as $\propto
\exp(-z^2/H^2)$ becomes negligible compared to that of the plume, and
so the hot gas of the plume can expand both vertically and
horizontally as a free expansion. With this $l$ we get
\begin{equation}
\frac{R_{shk}}{R_{\odot}}
\simeq \frac{1}{3}\frac {r}{H}
\label{eqn3.26} \mbox{.}
\end{equation} Using expression ~(\ref{H_rab}) for $H/r$ at $r \le
r_{ab}$ we obtain
\begin{equation} R_{shk} \simeq 0.24 H
\left(\frac{\alpha_{ss}}{0.01}\right)^{2/21}
\left(\frac{l_E}{0.1}\right)^{-26/21}
\left(\frac{\epsilon}{0.1}\right)^{26/21}
M_8^{-19/21}\frac{r}{r_{ab}}
\left(1-\sqrt{\frac{3r_g}{r}}\right)^{-2}
\label{eqn3.27}\mbox{,}
\end{equation} Thus a plume, starting from a size,
$R_{shk} < H$ will expand to a size $\simeq 2 H$ both vertically and
horizontally, thus producing a near spherical bubble with radius
$r_p=H$ above the disk. Post shock expansion  will increase the
estimate of $R_{shk}$ somewhat. For simplicity we will use
$R_{shk}=H/2$ for estimates of the toroidal flux entrained in the
plumes in paper~\uppercase{ii}. This is our standard plume.

Finally we note that the rise and fall time of this plume should be
the half orbit time, corresponding to a ballistic trajectory above
and back to the surface of the disk. Hence,   $t_{plume}
\simeq \pi / \Omega$ or  a plume rotation angle of $\pi$ radians. We
next consider the twisting of the plume leading to its effective
helicity.

\subsection{The Untwisting or Helicity Generation by the Plume}
\label{subsec_4.1}

Thus the   plume should expand to several times its original  radius
by the time it reaches the height of the order $2H$. The
corresponding increase in the moment of inertia of the plume and  the
conservation of the angular momentum of the plume causes the plume to
rotate slower relative to the inertial frame \citep{beckley03,
mestel99, colgate99}. From the viewpoint of the observer in the frame
corotating with the Keplerian flow at the  radius of the disk  of the
plume, this means that the plume rotates in the direction opposite to
the Keplerian rotation with an angular velocity equal to some
fraction of the local Keplerian angular velocity depending upon the
radial expansion ratio. Since the expansion of the plume  will not
be infinite in the rise and fall time of
$\pi$ radians of  Keplerian rotation of the disk, we expect that the
average of the plume rotation will be correspondingly less, or
$\Delta \phi < \pi$ or $\sim \pi/2$ radians. Any force or frictional
drag that resists this rotation will be countered by the Coriolis
force. Finally we note that kinetic helicity is proportional to
\begin{equation} h = {\bf v} \cdot ({\bf  \nabla}
\times {\bf v})
\label{eqn3.28} \mbox{.}
\end{equation} For the dynamo one requires one additional dynamic
property of the plumes. This is, that the total  rotation angle must
be finite and preferably
$\simeq
\pi/2$ radians, otherwise a larger angle or after many turns  the
vector of the entrained magnetic field would average to a small value
and consequently the  dynamo growth rate would be correspondingly
small.  This property of finite rotation, $\Delta
\phi
\sim \pi/2$ radians,  is  a fundamental property of plumes produced
above a Keplerian disk.

\section{Summary}
\label{sec_5}

Thus we have derived the  approximate properties of an accretion
disk around a massive black  hole, the high probability of star-disk
collisions, the  three necessary  properties of the resulting plumes
all necessary  for a robust dynamo. What is missing from this
description is the  necessary electrical properties of the medium. 
However, since the required conductivity is so closely related to the
mechanism of the dynamo itself, we leave it to the following paper~II
\citep{pariev06}, a discussion of this remaining property of the
hydrodynamic accretion disk flows necessary for a robust accretion
disk dynamo. With this exception we feel confident that an accretion
disk forming a CMBH with its associated star disk collisions is
nearly ideal for forming a robust feed-back-limited dynamo and thus,
potentially converting a major fraction of the gravitational free
energy of massive black hole formation into magnetic energy.

\acknowledgements

VP is pleased to thank Richard Lovelace and Eric Blackman for helpful
discussions. Eric Blackman is thanked again for his support during the
late stages of this work. SC particularly recognizes Hui Li of LANL
for support through the Director funded Research on the Magnetized
Universe and New Mexico Tech for support of the plume rotation
experiments as well as the dynamo experiment.  The facilities and
interactions of Aspen Center for Physics during two summer visits by
VP and more by SAC are  gratefully acknowledged. This work has been
supported by the U.S. Department of Energy through the  LDRD program
at Los Alamos National Laboratory. VP also acknowledges partial
support by DOE grant DE-FG02-00ER54600 and by the Center for Magnetic 
Self-Organization in Laboratory and Astrophysical Plasmas at the University
of Wisconsin-Madison.

\appendix

\section{Parameters of Shakura--Sunyaev Disk}
\label{appendix_A}

In the subsequent estimates of the disk physical parameters we will
keep  the radius of the disk $r$, where the star-disk collisions happen,
Shakura--Sunyaev viscosity parameter
$\alpha_{ss}$,  ratio of the disk luminosity  to the Eddington
luminosity
$l_E$, fraction
$\epsilon$ of  the rest mass accretion flux
${\dot M} c^2$, which is radiated away,  as parameters. We will
assume them to be  within an order of magnitude from their typical
values of importance for the dynamo problem, which are the following
\begin{equation}
\alpha_{ss}=0.01\mbox{,} \quad l_E=0.1\mbox{,}
\quad \epsilon=0.1\mbox{,}
\quad r=10^{-2}\,\mbox{pc}\label{typical}\mbox{.}
\end{equation} The flux of the stars through the disk,
$nv/4$, peaks at the radii inside
$r=10^{-2}\,\mbox{pc}$ (see section~\ref{subsec_stdiskrate}),
therefore we need to know the physics of the accretion disk at
$r\sim 10^{-2}\,\mbox{pc}$. Below, we will define the gravitational
radius as
$r_g=2GM/c^2 = 3.0\cdot 10^{13}M_8\,\mbox{cm} = 9.5\cdot
10^{-6}M_8\,\mbox{pc}$.  All formulae for the structure of
Shakura--Sunyaev disk are written for an arbitrary value of the black
hole mass $M=10^8 M_8 M_{\odot}$.  However, we will consider only
$M=10^8\,M_{\odot}$ whenever we invoke the  model for the star
distribution in the central cluster,  because the best available
model of the central star cluster was calculated for the $M=10^8
M_{\odot}$ (section~\ref{subsec_kinem}). Finally, the accuracy of
expressions for the disk parameters is only one significant figure in
all cases, and we keep two or even three figures only to avoid
introducing additional round off errors, when using our  expressions.
Similarly, one should not be concerned about small jumps  of values
across the boundaries with different physical approximations: 
a more elaborate treatment is needed to find exact matching solutions
there, although the physical principles are unchanged.

We use formulae from the
\citet{shakura73} article to obtain estimate of the state of the 
accretion disk. We assume the Schwarzschild black hole with the
inner edge  of the disk being at $3 r_g$. However, since we consider
star-disk  collisions happening at $\sim 10^3 r_g$, general
relativistic corrections are only at a level less than few per cents
and do not matter for our approximate treatment of star-disk
collision hydrodynamics. All expressions for disk quantities below
were also verified in later textbooks by
\citet{shapiro83} and
\citet{krolik99}.

The inner part of the disk (part (a) as in
\citet{shakura73})  is radiation dominated and the opacity is
dominated by Thomson scattering. In the next zone (part (b)) the
opacity is still Thomson, while the gas pressure exceeds radiation
pressure. In the outer most zone (part (c)) the opacity becomes
dominated by free-free and bound-free transitions. The boundary
between parts~(a) and~(b) $r_{ab}$ is given by an expression
\begin{equation} r_{ab}= 236 r_g
\left(\frac{\alpha_{ss}}{0.01}\right)^{2/21}
\left(\frac{M}{10^8\, M_{\odot}}\right)^{2/21}\left(\frac{l_E}{0.1}
\right)^{16/21}
\left(\frac{\epsilon}{0.1}\right)^{-16/21}
\label{rab}\mbox{.}
\end{equation} The boundary between parts (b) and (c)
$r_{bc}$ is given by the following expression
\begin{equation}  r_{bc}=3.4\cdot 10^3 r_g
\left(\frac{l_E}{0.1}\right)^{2/3}
\left(\frac{\epsilon} {0.1}\right)^{-2/3}
\label{rbc}\mbox{.}
\end{equation} One can see that, generally,
$r_{bc}>10^{-2}\,\mbox{pc}$. Therefore, we may consider zones 
(a) and (b) only, for our purpose of addressing star-disk
collisions.

First, we will list parameters following from solving for the
vertically averaged radial distributions of physical parameters
inside the zone~(a). The surface density is
\begin{eqnarray}
\Sigma & = & 407\,\mbox{g cm}^{-2}\,
\frac{0.01}{\alpha_{ss}}\left(\frac{l_E}
{0.1}\right)^{-1}\left(\frac{\epsilon}{0.1}
\right)\left(\frac{r
c^2}{GM}\right)^{3/2}\left(1-\sqrt{\frac{3r_g}{r}}
\right)^{-1}
    \nonumber \\ & = & 4.2\cdot 10^6\,\mbox{g cm}^{-2}\,
\left(\frac{\alpha_{ss}}{0.01}\right)^{-6/7}
\left(\frac{l_E}{0.1}
\frac{0.1}{\epsilon}\right)^{1/7} M_8^{1/7}
\left(\frac{r}{r_{ab}}\right)^{3/2}
\label{Sigma_a}\mbox{.}
\end{eqnarray} The half thickness of the disk is
\begin{equation} H = 2.6
\cdot 10^{13}\,\mbox{cm}\>\frac{l_E}{0.1}
\left(\frac{\epsilon}{0.1}
\right)^{-1}M_8\left(1-\sqrt{\frac{3r_g}{r}}
\right)\label{H_a}\mbox{.}
\end{equation} This $H$ depends upon the radius only via general
relativistic corrections. So, the disk has asymptotically constant
thickness for values of
$r\gg r_g$ \citep{shakura73, krolik99}. 
Moreover, $H$ does not depend on $\alpha_{ss}$ in zone (a)
and so $H$ is also independent on the mechanizm of angular 
momentum transport. 
The corresponding density is
\begin{eqnarray} &&
\rho=\frac{\Sigma}{2H}=7.5\cdot 10^{-7}\,\mbox{g cm}^{-3}\>
\frac{0.01}{\alpha_{ss}}
\left(\frac{l_E}{0.1}\right)^{-2}
\left(\frac{\epsilon}{0.1}\right)^2
\times
\nonumber \\  &&
\left(\frac{r}{10^{-2}\,\mbox{pc}}
\right)^{3/2} M_8^{-5/2}
\left(1-\sqrt{\frac{3r_g}{r}}\right)^{-2}
\label{rho}\mbox{.}
\end{eqnarray} At the luminosity of an AGN
\begin{equation} L= 1.3\cdot 10^{45}\left(\frac{l_E}{0.1}\right)
M_8\,
\mbox{erg s}^{-1}\label{L}\mbox{,}
\end{equation} the  mass flux is
\begin{equation} {\dot M}=0.23\,M_{\odot}\mbox{yr}^{-1}
\left(\frac{0.1}{\epsilon}\right)
\left(\frac{l_E}{0.1}\right)M_8 = 1.4\cdot 10^{25}\,\mbox{g s}^{-1}
\left(\frac{0.1}{\epsilon}\right)
\left(\frac{l_E}{0.1}\right)M_8
\label{Mdot}\mbox{.}
\end{equation}

The energy emitted from the unit surface of the one side of the  disk
per unit time is
\begin{eqnarray} && Q=\frac{3}{8\pi}{\dot
M}\frac{GM}{R^3}\left(1-\sqrt{\frac{3r_g}{r}}
\right)=7.6\cdot 10^8\,\mbox{erg cm}^{-2}\mbox{s}^{-1}\,
\left(\frac{0.1}{\epsilon}\right)
\left(\frac{l_E}{0.1}\right)
\times
\nonumber \\ &&
\left(\frac{r}{10^{-2}\,\mbox{pc}}\right)^{-3} M_8^2
\left(1-\sqrt{\frac{3r_g}{r}}\right)
\label{Q}\mbox{.}
\end{eqnarray} The effective temperature near the surface of the disk
is
\begin{eqnarray} && Q=\frac{ac}{4}T_s^4\mbox{,}
\quad T_s=1900\,\mbox{K}
\left(\frac{0.1}{\epsilon}\right)^{1/4}
\left(\frac{l_E}{0.1}\right)^{1/4}
\times
\nonumber \\ &&
\left(\frac{r}{10^{-2}\,\mbox{pc}}
\right)^{-3/4}M_8^{1/2}
\left(1-\sqrt{\frac{3r_g}{r}}\right)^{1/4}
\label{Tsurf}\mbox{.}
\end{eqnarray} The article by
\citet{shakura73} contains the solution of the radiative transport
equation in the vertical direction of an optically thick disk  with
assumed local thermodynamic equilibrium for each $z$ in the disk.
Volume emission due to viscous heating is included in the solution.
Using this solution with the Thomson opacity
$\kappa_T=0.4\,\mbox{cm}^2
\mbox{g}^{-1}$  one obtains (section~2a of
\citealt{shakura73})  the temperature at the midplane of the disk
\begin{equation} T_{mpd}^4=T_s^4\left(1+\frac{3}{16}\kappa_T
\Sigma\right)\label{Tmpdexact}
\mbox{.}
\end{equation} Since the disk is very opaque for Thomson scattering
one can neglect 1 in  the expression~(\ref{Tmpdexact}) and obtains
\begin{eqnarray} && T_{mpd}=T_s\,\cdot
41.3\left(\frac{0.01}{\alpha_{ss}}\right)^{1/4}
\left(\frac{l_E}{0.1}\right)^{-1/4}
\left(\frac{\epsilon}{0.1}\right)^{1/4}
\times
\nonumber \\ &&
\left(\frac{r}{10^{-2}\,\mbox{pc}}\right)^{3/8} M_8^{-3/8}
\left(1-\sqrt{\frac{3r_g}{r}}\right)^{-1/4}
\label{T1}\mbox{.}
\end{eqnarray} Using expression~(\ref{Tsurf}) for the effective
surface temperature
$T_s$ and substituting for
$T_s$ in the equation~(\ref{T1}) one obtains
\begin{equation} T_{mpd}=7.9\cdot
10^4\,\mbox{K}\,\left(\frac{0.01}{\alpha_{ss}}
\right)^{1/4}
\left(\frac{r}{10^{-2}\,\mbox{pc}}\right)^{-3/8} M_8^{1/8}
\label{Tmpd}\mbox{.}
\end{equation} Note that the terms describing the dependence on the
accretion rate cancel out as well as general relativistic correction
term. Therefore,
$T_{mpd}$ in the inner parts of accretion disk does not depend on 
the accretion rate, but is determined by the mass of the central
black hole and viscosity parameter
$\alpha_{ss}$. Radiation pressure in the midplane of the disk in the
zone~(a) is
\begin{equation} P_r=\frac{1}{3}a T_{mpd}^4 = 1.07\cdot
10^5\,\mbox{erg cm}^{-3}
\frac{0.01}{\alpha_{ss}} M_8^{1/2}
\left(\frac{r}{10^{-2}\,\mbox{pc}}
\right)^{-3/2}
\label{Pr_a}\mbox{.}
\end{equation}

By integrating surface density~(\ref{Sigma_a}) one can obtain the
total mass of the  disk inside radius $r$ (assuming $r\gg r_g$)
\begin{equation} M_{disk}=10^8\,M_{\odot}M_8\left(\frac{r}
{r_{sg}}\right)^{7/2}
\label{Mdisk}\mbox{,}
\end{equation} where the radius
$r_{sg}$, inside of which the mass of the disk is equal to the mass
of the black hole, is given  by
\begin{equation} r_{sg}=1400 r_g\cdot M_8^{-2/7}
\left(\frac{\alpha_{ss}}{0.01}\right)^{2/7}
\left(\frac{l_E}{0.1}\right)^{2/7}
\left(\frac{\epsilon}{0.1}\right)^{-2/7}
\label{rsg}\mbox{.}
\end{equation} Since the total mass of the disk grows very rapidly
with the radius
$r$, the gravitational potential of the disk would dominate the
gravitational potential of the black hole for $r>r_{sg}$. As follows
from comparing  expressions~(\ref{rab}) and~(\ref{rsg})
$r_{sg}>r_{ab}$ in general. More exactly, the condition
$r_{sg}>r_{ab}$ reduces to
\begin{equation} M_8^{-8/21}\left(\frac{\alpha_{ss}}{0.01}
\right)^{4/21}\left(\frac{l_E}{0.1}
\right)^{-10/21}\left(\frac{\epsilon}{0.1}
\right)^{10/21} > 0.165
\label{rsg<rab}\mbox{.}
\end{equation} The dependence of the left hand side of
equation~(\ref{rsg<rab}) on the  black hole mass
$M_8$ and viscosity parameter
$\alpha_{ss}$ is weak. One also expects the efficiency of radiation
$\epsilon$ to be within the order of magnitude from the value
$0.1$. The largest variations are expected for the luminosity
$l_E$. However, even for $l_E=1$, still
$r_{ab}<r_{sg}$. Generally, the mass of the inner zone of the disk is
small compared to the mass of the black hole. Below we assume
$r_{ab}<r_{sg}$ to be  always satisfied. The total mass of the inner
part of the disk enclosed inside
$r<r_{ab}$ is
\begin{equation} M(r_{ab})=1.83\cdot
10^5\,M_{\odot}\left(\frac{\alpha_{ss}}{0.01}
\right)^{-2/3} M_8^{7/3}\left(\frac{l_E}{0.1}\right)^{5/3}
\left(\frac{\epsilon}{0.1}\right)^{-5/3}
\label{Ma}\mbox{,}
\end{equation} which is, generally, much smaller than the mass $10^8
M_8 M_{\odot}$ of the  black hole.

In the zone (b) the surface density is given by
\begin{eqnarray}
\Sigma & = & 1.75\cdot 10^8\,\mbox{g cm}^{-2}\,
\left(\frac{\alpha_{ss}}{0.01}\right)^{-4/5}
\left(\frac{l_E}{0.1}\right)^{3/5}
\left(\frac{\epsilon}{0.1}\right)^{-3/5} M_8^{1/5}\left(\frac{r
c^2}{GM}\right)^{-3/5}
\nonumber
\\ & = & 4.4\cdot 10^6 \,\mbox{g cm}^{-2}\,
\left(\frac{r}{r_{ab}}\right)^{-3/5}
\left(\frac{\alpha_{ss}}{0.01}\right)^{-6/7} M_8^{1/7}
\left(\frac{l_E}{0.1}\frac{0.1}{\epsilon}
\right)^{1/7}
\label{Sigma_b}\mbox{.}
\end{eqnarray} The integral of this surface density from
$r_{ab}$ to any given $r$ gives  the mass enclosed between $r_{ab}$
and $r$ as
\begin{eqnarray} && M_b(r)=85\,M_{\odot}
\left[\left(\frac{r}{M}\right)^{7/5}-
\left(\frac{r_{ab}}{M}\right)^{7/5}
\right]
\left(\frac{\alpha_{ss}}{0.01}\right)^{-4/5}
\times
\nonumber
\\ &&
\left(\frac{l_E}{0.1}\right)^{3/5}
\left(\frac{\epsilon}{0.1}\right)^{-3/5}
M_8^{11/5}\label{M_b}\mbox{.}
\end{eqnarray} Now we can estimate the value of
$r=r_{sg}$ such that
$M_b(r_{sg})=10^8\,M_8 M_{\odot}$ (neglecting the contribution from
the part (a) of the disk). Neglecting
$1$ compared to the ratio
$r_{sg}/r_{ab}\gg 1$, one obtains
\begin{equation}
\frac{r_{sg}}{r_{ab}}\approx 46\, M_8^{-20/21}
\left(\frac{\alpha_{ss}}{0.01}\right)^{10/21}
\left(\frac{l_E}{0.1}\right)^{-25/21}
\left(\frac{\epsilon}{0.1}\right)^{25/21}
\label{rsgb}\mbox{.}
\end{equation} The logarithmic width of the zone (b), i.e. the ratio
$r_{bc}/r_{ab}$, is  given by
\begin{equation}
\frac{r_{bc}}{r_{ab}}=14.4\left(\frac
{\alpha_{ss}}{0.01}\right)^{-2/21} M_8^{-2/21}\left(\frac{l_E}{0.1}
\right)^{-2/21}
\left(\frac{\epsilon}{0.1}\right)^{2/21}
\label{univconstant}\mbox{,}
\end{equation} i.e. almost a constant, depending on all parameters
of the disk and the black hole very weakly. Depending upon
parameters, $r_{sg}$ maybe inside or outside the $r_{bc}$, however,
as we show next, the disk in part~(b) is unstable to fragmentation
caused by self gravity, which makes the question on whether the exact
position of
$r_{sg}$ is with respect to $r_{bc}$ unimportant. The expressions for
radiation flux
$Q$ and surface temperature of the disk
$T_s$ remain the same as in the part~(a) of the disk, namely given by
the expressions~(\ref{Q}) and~(\ref{Tsurf}). For the temperature at
the midplane of the disk one can obtain from formula~(\ref{Tmpdexact})
\begin{equation} T_{mpd}=3.5\cdot
10^7\,\mbox{K}\,\left(\frac{\alpha_{ss}}{0.01}
\right)^{-1/5}
\left(\frac{0.1}{\epsilon}\frac{l_E}{0.1}
\right)^{2/5}
\left(\frac{r c^2}{GM}\right)^{-9/10}
M_8^{-1/5}\label{Tmpd_b}\mbox{.}
\end{equation}

The characteristic thickness of the disk is given by
\begin{equation} H = 2.75\cdot
10^{10}\,\mbox{cm}\,\left(\frac{\alpha_{ss}} {0.01}\right)^{-1/10}
\left(\frac{0.1}{\epsilon}\frac{l_E}{0.1}
\right)^{1/5}M_8^{9/10}
\left(\frac{r c^2}{GM}\right)^{21/20}\label{H_b}\mbox{.}
\end{equation} Then, from expressions~(\ref{Sigma_b})
and~(\ref{H_b}), one can obtain the vertically averaged density in
the zone~(b) as
\begin{equation}
\rho = \frac{\Sigma}{2 H} = 3.2 \cdot 10^{-3}\,\mbox{g cm}^{-3}\,
\left(\frac{\alpha_{ss}}{0.01}\right)^{-7/10}
\left(\frac{l_E}{0.1}
\frac{0.1}{\epsilon}\right)^{2/5} M_8^{-7/10}
\left(\frac{r c^2}{GM}\right)^{-33/20}
\label{rho_zoneb}\mbox{.}
\end{equation} The corresponding values of the radiation pressure
$\displaystyle P_r=\frac{1}{3}aT_{mpd}^4$ and the gas pressure
$P_g=2nkT_{mpd}$ at the midplane are
\begin{eqnarray} && P_r= 3.8\cdot 10^{15}\,\mbox{erg cm}^{-3}
\left(\frac{\alpha_{ss}}{0.01}\right)^{-4/5}
\left(\frac{0.1}{\epsilon}
\frac{l_E}{0.1}\right)^{8/5}
\times
\nonumber
\\  &&
\left(\frac{r c^2}{GM}\right)^{-18/5} M_8^{-4/5}
\left(1-\sqrt{\frac{3 r_g}{r}}\right)^{8/5}\label{Pr_b}\mbox{,}
\\ && P_g= 1.76\cdot 10^{13}\,\mbox{erg cm}^{-3}
\left(\frac{\alpha_{ss}}{0.01}\right)^{-9/10}
\left(\frac{0.1}{\epsilon}\frac{l_E}{0.1}
\right)^{4/5}
\times
\nonumber \\ && \left(\frac{r c^2}{GM}\right)^{-51/20} M_8^{-9/10}
\left(1-\sqrt{\frac{3 r_g}{r}}\right)^{4/5}\label{Pg_b}\mbox{.}
\end{eqnarray} Calculating the ratio of $P_r$ to
$P_g$ one can recover the expression~(\ref{rab}) for the radius, when
$P_r=P_g$.

When the disk becomes self gravitating, it may become subject to
gravitational instability. Let us check that by calculating  Toomre
parameter
$\displaystyle \mbox{To}=\frac{\varkappa c_s}{\pi G \Sigma}$ (e.g.,
\citealt{binney94}). Epicyclic frequency
$\varkappa$ is equal to its value for the point mass $M$ located at
the position of the black  hole
$\varkappa=\Omega_K=(GM)^{1/2}/r^{3/2}$, since the mass of the disk
is  small compared to the mass of the black hole.  Sound speed is
equal to
$\displaystyle c_s^2=\frac{kT_{mpd}}{m_p}$ in zone~(b) and 
$\displaystyle c_s^2=\frac{P_r}{\rho}$ in the zone~(a) (a coefficient
close to
$1$ is neglected). Substituting appropriate  expressions we obtain
for the sound speed in zone~(b)
\begin{eqnarray} && c_s=5.37\cdot 10^7\,\mbox{cm
s}^{-1}\,\left(\frac{\alpha_{ss}}{0.01}
\right)^{-1/10}
\left(\frac{0.1}{\epsilon}\frac{l_E}{0.1}
\right)^{1/5}
\times
\nonumber \\ && \left(\frac{r c^2}{GM}\right)^{-9/20} M_8^{-1/10}
\left(1-\sqrt{\frac{3 r_g}{r}}\right)^{1/5}
\label{cs_b}\mbox{,}
\end{eqnarray} in zone~(a)
\begin{equation} c_s=3.5\cdot 10^{10}\,
\mbox{cm s}^{-1}\,\frac{l_E}{0.1}
\left(\frac{\epsilon}{0.1}\right)^{-1}
\left(\frac{r c^2}{GM}\right)^{-3/2}
\left(1-\sqrt{\frac{3 r_g}{r}}\right)\label{cs_a}\mbox{.}
\end{equation}  The Toomre parameter becomes in zone~(a)
\begin{eqnarray} &&
\mbox{To}=8.33\cdot 10^{11}\,\frac{\alpha_{ss}}{0.01}
\left(\frac{l_E}{0.1}\right)^2
\left(\frac{\epsilon}{0.1}\right)^{-2}
\left(\frac{r c^2}{GM}\right)^{-9/2} M_8^{-1}\left(1-\sqrt{\frac{3
r_g}{r}}\right)^2
\nonumber \\ && = 0.77\,\left(\frac{\alpha_{ss}}{0.01}
\right)^{4/7} M_8^{-10/7}
\left(\frac{l_E}{0.1}\right)^{-10/7}
\left(\frac{\epsilon}{0.1}\right)^{10/7}
\left(\frac{r}{r_{ab}}\right)^{-9/2}
\label{Q_a}
\end{eqnarray} and in the zone~(b)
\begin{eqnarray} && \mbox{To}= 2.97\cdot 10^3\,
\left(\frac{\alpha_{ss}}{0.01}\right)^{7/10}
\left(\frac{0.1}{\epsilon}
\frac{l_E}{0.1}\right)^{-2/5} M_8^{-13/10}
\left(\frac{r c^2}{GM}\right)^{-27/20}
\nonumber \\ && = 0.73
\left(\frac{r}{r_{ab}}\right)^{-27/20} M_8^{-10/7}
\left(\frac{\alpha_{ss}}{0.01}\right)^{4/7}
\left(\frac{l_E}{0.1}\frac{0.1}{\epsilon}
\right)^{-10/7}\label{Q_b}
\\ && = 9.7\cdot 10^{-2}\left(\frac{\alpha_{ss}}{0.01}
\right)^{7/10}
\left(\frac{l_E}{0.1}\frac{0.1}{\epsilon}
\right)^{-2/5} M_8^{1/20}
\left(\frac{r}{0.01\,\mbox{pc}}\right)^{-27/20}
\mbox{.}
\nonumber
\end{eqnarray} Gravitational instability  develops if
$\mbox{To}<1$. One can see from expressions~(\ref{Q_a})
and~(\ref{Q_b}) that $\mbox{To}$ strongly declines with increasing
the radius. The disk has a well  defined outer radius of
gravitational stability $r_T$ such that
$\mbox{To}(r_T)=1$.  For our fiducial parameters,
$r_T$ is close to the $r_{ab}$. At the outer edge of the zone~(b) one
has
\begin{equation}
\mbox{To}(r_{bc})=2.0\cdot 10^{-2}\left(\frac{\alpha_{ss}}{0.01}
\right)^{7/10} M_8^{-13/10}\left(\frac{l_E}{0.1}\frac{0.1}
{\epsilon}\right)^{-13/10}
\label{Qr_bc}\mbox{.}
\end{equation} Large values of
$\alpha_{ss}$, small masses of the central black hole, and low
accretion rates cause the
$\mbox{To}$ to increase and can cause the radius $r_T$ to become
larger than
$r_{ab}$. As follows from expression~(\ref{Q_a}) the value for
$r_T$ (when $r_T<r_{ab}$) is
\begin{eqnarray} && r_T\approx r_{ab}\,\left(\frac{\alpha_{ss}}{0.01}
\right)^{8/63} M_8^{-20/63}
\left(\frac{l_E}{0.1}\right)^{-20/63}
\left(\frac{\epsilon}{0.1}\right)^{20/63}
\left(1-\sqrt{\frac{3 r_g}{r}}\right)^{4/9}\nonumber\\ && = 218 r_g
M_8^{-2/9}
\left(\frac{\alpha_{ss}}{0.01}\right)^{2/9}
\left(\frac{0.1}{\epsilon}\frac{l_E}{0.1}
\right)^{4/9}\label{r_Q}\mbox{.}
\end{eqnarray} Assuming the range of parameters
$1>\alpha_{ss}>10^{-3}$,
$10^{-2}<M_8<10^2$,
$10^{-3}<l_E<1$, and $\epsilon \approx 0.1$ the lowest possible
location of
$r_T$ will be at $\approx 6 r_g$, i.e. in the vicinity of the inner
edge of the disk, where the Toomre stability criterion is not directly
applicable. On the other side, the stable region  of the disk can
extend over the whole of zone~(b) and into the outermost zone~(c) as
well. At the radius of
$r=0.01\,\mbox{pc}$ and $M_8=1$,  which corresponds to the width of
the broad line region, the disk would be unstable for the fiducial
set of parameters. However, at lower values of accretion rates $l_E
\leq 0.01$, which are expected in the case of relatively low activity
in Seyfert  galaxies, and larger values of
$\alpha_{ss}
\geq 0.1$ the stable part of the  disk will include
$0.01\,\mbox{pc}$.

\newpage

\begin{figure}
\epsscale{0.5}
\plotone{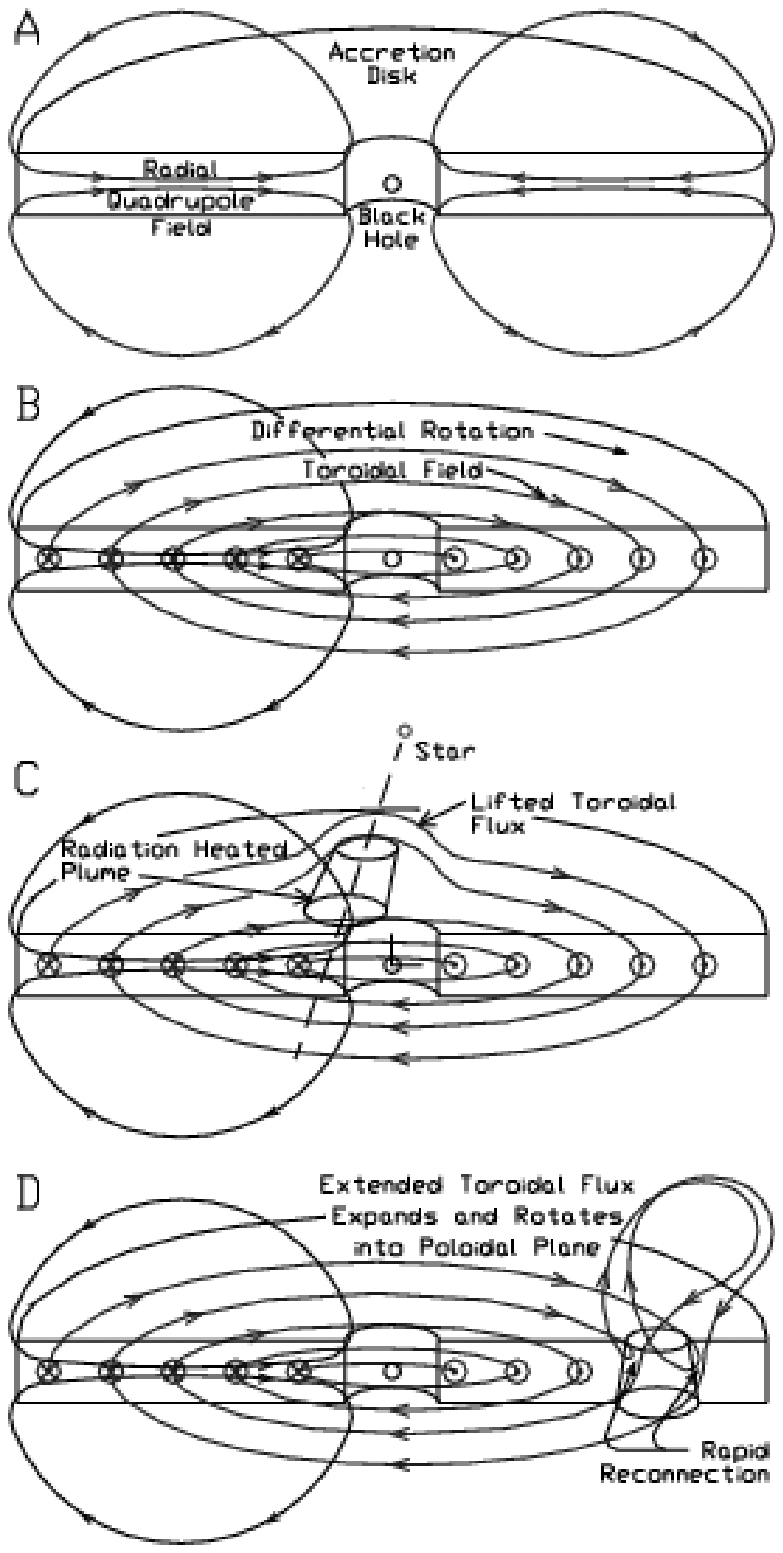}
\caption {The $\alpha-\Omega$ dynamo in a galactic black hole
accretion disk. The radial component of the poloidal quadrupole
field within the disk (A) is sheared by the differential rotation
within the disk, developing a stronger toroidal component (B).  As a
star passes through the disk it heats by shock and by radiation a
fraction of the matter of the disk, which expands vertically and
lifts a fraction of the toroidal flux within an expanding plume (C).
Due to the conservation of angular momentum, the expanding plume and
embedded flux rotate
$\sim \pi/2$ radians before the matter in the plume and embedded
flux  falls back to the disk (D).  Reconnection allows the new
poloidal flux to merge with and augment the original poloidal flux
(D).}
\label{fig1a}
\end{figure}

\end{document}